**Full Title:**

# Factorial clinical trials in the presence and absence of plausible statistical interactions between treatment factors. A historical review of the methodological literature.

**Short Title:**

Statistical interactions in factorial clinical trials

**Authors:**

Rebecca E.A. Walwyn (University of Leeds, UK) ORCID ID 0000-0001-9120-1438

Pritaporn Kingkaew (HITAP, Thailand)

Alan A. Montgomery (University of Nottingham, UK)

Alexandra Wright-Hughes (University of Leeds, UK)

Steven G. Gilmour (King's College London, UK)

**Contact Information for Corresponding Authors:**

Rebecca Walwyn, Leeds Institute for Clinical Trials Research, University of Leeds, Leeds, United Kingdom, LS2 9JT.

Email: R.E.A.Walwyn@leeds.ac.uk

**Acknowledgments**

None.

**Statements and Declarations**

The authors have no conflicts of interest to declare. Ethical approval was obtained for Pritaporn's PhD. No additional ethical approval was required for this article. No new data were generated by this work.

**Funding Statement**

Rebecca Walwyn was funded by a National Institute for Health Research (NIHR) Advanced Fellowship (ref: NIHR301709). Pritaporn Kingkaew's PhD was funded by HITAP, Thailand. Other authors received no financial support.

## Abstract

Factorial trials can be conducted when statistical interactions between two or more treatment factors are not anticipated, but also when they are. A *k*-in-1 factorial trial answers *k* single-factor questions with the same number of units as one parallel-group trial if there are no interactions. Literature on *k*-in-1 factorial trials has dominated trialists understanding of factorial trials in the UK. However, factorial experiments originated from the Design of Experiments field to enable interactions to be robustly estimated. Seemingly conflicting guidance from these literatures poses a source of confusion and misunderstanding for trialists. We bring these literatures together to provide clarity on the arguments that have been used to recommend use of factorial trials in the presence and absence of interactions. We outline motivating examples. We summarise the rationales for using factorial trials, the treatment contrasts of interest, and the properties of their estimators, for the two schools of thought. We describe the debate, going back to 1935, and use an empirical example to illustrate the impact of different analysis approaches. We conclude that it is vital that trialists carefully and clearly specify their objectives. Estimands of interest and treatment contrasts follow, with properties of estimators dictated by this choice.

## 1. INTRODUCTION

In a factorial trial, individuals or clusters of individuals are randomised to one combination of the levels of two or more treatment factors (that is, a "treatment"). The simplest case is an individually-randomised 2×2 factorial trial with two treatment factors, $A$ and $B$, each with two levels, 1 and 2. Individuals are randomised to receive $A_1$ and $B_1$, $A_2$ and $B_1$, $A_1$ and $B_2$, or $A_2$ and $B_2$.[1] Here, individuals are experimental units, and there are four "experimental conditions" or "treatments". The highest level of each treatment factor may be a "whole" or entire intervention (e.g., a drug where interest lies in that drug in isolation), or a version of a component of a "complex" intervention[2,3,4] (e.g., support where interest lies in the support as part of an overall intervention package). Each treatment factor may have just two levels (present or absent, enhanced or standard, high or low), or three or more levels. More complex factorial trials entail three or more treatment factors, at two or more levels, or a carefully chosen fraction of all the treatment factor combinations, allowing estimation of treatment effects of interest.[5] While these design options are accepted in a health and social care context, there exist two schools of thought that are a source of confusion and misunderstanding for clinical trialists.

In the first school of thought, originating in the UK, factorial trials are undertaken only in the assumed absence of statistical interactions between treatment factors.[1,6] This school of thought can be traced back to 1978, when Peto[7] suggested that factorial trials can be used to evaluate multiple whole interventions, without increasing the sample size from that required for one parallel-group trial. Peto[7] claimed that a 2×2 factorial trial can provide "two independent answers for the price of one" (p. 35). A strand of literature has followed on this specific use of a factorial trial, which has largely dominated understanding of factorial trials by trialists in the UK. For clarity, and consistent with Kahan *et al*,[1] we will refer to these factorial trials as $k$-in-1 factorial trials. The first factorial clinical trial predates 1978 though. It has been credited to Chalmers *et al*[8] in 1955 (see[9,10]). However, Doll[11] refers to two early precedents,[12,13] both using alternation to assign patients to treatments. To our knowledge, the first randomised factorial clinical trial was published in the *Lancet* in 1952.[14] Even so, factorial clinical trials were rarely conducted until the 1980s.[15,16]

In the second school of thought, originating from the field of mathematical statistics known as the Design of Experiments, factorial trials are recommended in the presence and absence of statistical interactions between treatment factors.[17,18,19] This school of thought began with Fisher[20] and Yates,[21] who referred to factorial experiments as "complex experiments".

Factorial experiments have an even longer history, dating at least to the 1840s at Rothamsted.[21] They were originally developed for use in agriculture, with the terms "factorial design" and "factorial experiment" dating back to Fisher[17] and then Yates[21] in 1935. Since, an extensive body of literature has developed, starting with Yates[22] in 1937, and culminating in introductory Design of Experiments textbooks (e.g.,[5,17,23–28]) including chapters on factorial experiments, and other textbooks being dedicated to factorial experiments.[29,30]

More recently, Collins *et al*[31] proposed that factorial trials can be used to optimise multi-component interventions in a research framework referred to as the Multiphase Optimisation Strategy (MOST). The strand of literature that followed is summarised in two books,[18,19] with a more recent overview given in.[32] Collins *et al*[18,19,31,32] were influenced by the engineering field, and especially by the Design of Experiments textbooks by Box, Hunter and Hunter[5] and Wu and Hamada.[28] For clarity, and consistent with MOST,[32] we will refer to this specific use of a factorial trial as a factorial optimisation trial. An early example was published in 2016 by Piper *et al*[33]. More recently still, Dasgupta *et al*[34] has placed factorial trials into the potential outcomes framework. Since then, a small strand of literature has developed in the causal inference field, largely unrecognised by trialists. While the MOST literature clearly fits into the second school, the causal inference literature may be viewed as straddling both schools of thought.

At first sight, the two schools of thought appear to conflict. Nevertheless, we suggest that the necessity for an absence of interactions only applies to how the sample size is calculated and to the choice of analysis model for $k$-in-1 factorial trials; it does not apply to factorial trials in general. The "estimands" of interest, the choice of treatment contrasts, the properties of their estimators, and the approach to decision making all differ depending on the objectives of the trial (for an overview, see Table 1). The aim of this paper is to provide some clarity on the arguments which have been used to recommend that factorial clinical trials are used either in the presence or absence of plausible interactions.

[Insert Table 1 about here]

Section 2 outlines an example of a $k$-in-1 factorial trial[35] (for the first school of thought), and a fractional factorial screening experiment,[36] a factorial optimisation trial,[37] and a factorial evaluation trial[38] (for the second school of thought) to highlight their differences. Section 3 outlines the rationale for each school of thought, definitions of treatment effects, properties of their estimators, and approaches to decision making. Section 4 summarises the debate on interactions. Section 5 presents an empirical example to illustrate the impact of the different approaches to factorial trials, and Section 6 ends with some conclusions.

## 2. MOTIVATING EXAMPLES

### 2.1 *k*-in-1 Factorial Trial

The Second International Study of Infarct Survival (ISIS-2)[35] was a 2-in-1 factorial trial. In total, 17187 patients with suspected acute myocardial infarction were randomised to either (i) placebo streptokinase ($A_1$) and placebo aspirin ($B_1$), (ii) streptokinase ($A_2$) and placebo aspirin ($B_1$), (iii) placebo streptokinase ($A_1$) and aspirin ($B_2$), or (iv) streptokinase ($A_2$) and aspirin ($B_2$) in a 2×2 factorial design. ISIS-2 also exemplifies a large, simple trial in the sense advocated by Yusuf *et al*.[39] The aim of the trial was to compare (1) streptokinase ($A_2$, $B_1$) to placebo streptokinase ($A_1$, $B_1$) both with placebo aspirin ($B_1$) on vascular mortality in (i) the first 5 weeks and (ii) entire period, and (2) aspirin ($A_1$, $B_2$) to placebo aspirin ($A_1$, $B_1$) both with placebo streptokinase ($A_1$) in the first 5 weeks. However, instead of including just three treatments (($A_2$, $B_1$), ($A_1$, $B_2$) and ($A_1$, $B_1$)), efficiency was gained by also including the fourth treatment ($A_2$, $B_2$) to allow all the patients to be included in estimating each of the treatment effects of primary interest. This was done by assuming no statistical interaction between the effects of streptokinase and aspirin and estimating the effect of streptokinase ($A_2$) compared to placebo streptokinase ($A_1$), for example, by comparing $A_2$ to $A_1$ both with placebo aspirin ($B_1$) and with aspirin ($B_2$) using a marginal effect. The no interaction assumption allowed the sample size required for the trial to be reduced compared to a three-arm parallel-group trial, and more questions to be answered compared to a two-arm parallel-group trial. Patients randomised to streptokinase and aspirin ($A_2, B_2$) were compared also to those randomised to placebo streptokinase and placebo aspirin ($A_1, B_1$) in a secondary analysis. Details of an interaction between the treatment factors are not given but it is stated that "effects on vascular deaths appeared to be additive" (p. 349) and that "benefits appeared to be largely independent of each other" (p. 357). This was in support of their assumption of no statistical interaction, not a finding of primary interest.

### 2.2 Fractional Factorial Screening Experiment

ENACT[36] was a $2^{6-1}$ fractional factorial screening experiment. In total, 638 clinicians, managers and audit staff were randomised to receive one of 32 combinations of the two levels of six specific modifications ($A$ to $F$) to feedback given alongside five national clinical audits. The modifications were effective comparators ($A$), multimodal feedback ($B$), specific actions ($C$), optional detail ($D$), patient voice ($E$) and cognitive load ($F$). A half-fraction (32 of 64 treatment factor combinations) was chosen for logistical reasons. The aim of the experiment was to identify the leading candidates for future real-world evaluation. The intention was to screen out clearly ineffective components of feedback under laboratory conditions. Sample size was calculated to give 90% power to detect small-to-moderate main effects (0.3 SDs) for each modification using a 2-sided 5% significance level in the presence of interactions between modifications. The focus was in estimating the effect of each modification averaged across all levels of the other modifications (the main effects), and the primary analysis included all main effects, two-way interactions and a prespecified half of the three-way interactions, recognising that the three-way interactions were aliased with other three-way interactions and higher order interactions were aliased with two-way interactions and main effects. A parsimonious statistical model was identified, determined by statistical significance of effects and a Pareto plot. Conclusions were directions for future research.

## 2.3 Factorial Optimisation Trial

Kingkaew *et al*[37] conducted a 2×2×2 factorial optimisation trial. In this trial, 1571 smokers were randomised to receive one of eight sets of text messages with behavioural change techniques to enhance the (i) capability, (ii) opportunity and (iii) motivation of people to stop smoking using the COM-B model of behaviour change.[40] There were two levels for each treatment factor. The aim of the trial was to better understand how the text messages could aid smoking cessation, while testing the COM-B model and predicting the optimal treatment. The sample size was calculated to detect minimally important main effects with 80% power with a 2-sided 5% significance level in the presence of interactions between the treatment factors. Primary objectives were to estimate the main effects. The interactions between treatment factors were of secondary interest. The primary analysis included all main effects, two-way interactions and the three-way interaction. The full model was reported. Conclusions were made about the importance of each text message component and the directions for future research.

## 2.4 Factorial Evaluation Trial

Each cluster-randomised trial in the AFFINITIE programme[38] was a 2×2 factorial evaluation trial. In each trial, about 130 NHS trusts were randomised to receive (i) usual audit reports ($A_1$) and usual support for staff ($B_1$), (ii) enhanced reports ($A_2$) and usual support ($B_1$), (iii) usual reports ($A_1$) and enhanced support ($B_2$), or (iv) enhanced reports ($A_2$) and enhanced support ($B_2$). Primary outcomes were measured on patients in trusts. Each trial aimed to evaluate two theoretically-informed feedback components embedded within a real-world National Clinical Audit programme of blood transfusion practice in UK hospitals, where each component might be given with either the usual or the enhanced version of the other component. As feedback is a complex intervention, a statistical interaction between the two components was considered plausible. The sample size was calculated to enable detection of minimally important main effects, allowing for an interaction, with 80% power using a 2-sided 5% significance level. Primary objectives were to estimate main effects of reports and supports and the interaction. Conclusions were made about the optimal treatment.

## 3. HISTORICAL REVIEW OF FACTORIAL CLINICAL TRIALS

### 3.1 Factorial Experiments

Yates[21] attributed the development of factorial experiments in agricultural research chiefly to Fisher, suggesting that following his[20] recommendation, factorial experiments were used "extensively" at Rothamsted "and to a lesser extent elsewhere" (p. 189). Fisher[17,20] gave his rationale for using a factorial experiment rather than a separate parallel-group experiment for each treatment factor (Fisher referred to the latter as single-factor experiments). Yates[21] then provided detail on "the methods in present use" (p. 189) (Yates[22] gives a complete description). Introductory Design of Experiments textbooks[5, 23–28] draw heavily on these early 20$^{th}$ century texts.[17,20,21,22] The MOST framework for clinical trials[18,19,31,32] draws heavily on two of these textbooks.[5,28]

Fisher[20] argued that the rationale for conducting factorial experiments is threefold. Firstly, single-factor comparisons can be made "with known and with, probably, high accuracy" (p. 93). Secondly, "many other comparisons" (by which he means interactions) can be made with "equal accuracy" (p. 93). Thirdly, and "most important", "the conclusions drawn from the single-factor comparisons will be given, by the variation of non-essential conditions, a very much wider inductive basis than could be obtained, by single question methods, without extensive repetitions of the experiment" (p. 93). Fisher[17] then expands his initial rationale. He believed experimenters usually have "no knowledge that any one [treatment] factor will exert its effects independently of all others that can be varied, or that its effects are…simply related to variations in these other factors" (p. 97). He continued, "The modifications possible to any complicated apparatus, machine, or industrial process must always be considered as potentially interacting with one another, and must be judged by the probable effects of such interactions" (p. 97). Of a mixture (that is, a complex intervention), Fisher[17] states, "the general question as to whether, of each ingredient, more or less can be added with advantage, can be settled by making up all combinations, using either more or less of each component" (p. 99). Fisher[17,20] thus suggested experimenters are interested in estimating "general" (i.e., "main" or "marginal") effects and interactions.

Fisher[17] said that efficiency is the benefit of a factorial experiment over a series of single-factor experiments, with "more knowledge and a higher degree of precision…obtainable by the same number of observations" (p. 98). He stated that efficiency comes from four sources. Firstly, "no light whatever [is given] on the possible interactions of the different ingredients" by a series of single-factor experiments (p. 100). Estimating interactions is only possible "by carrying out our test of A both in the presence and in the absence of B…[which] is what the factorial experiment is designed to do" (p. 101). Secondly, Fisher[17] argues "in addition to measuring the effects of…single ingredients with the same precision as though the whole of the experiment had been devoted to each of them", "interactions between these ingredients [are measured] with the same precision" (p. 104). Thirdly, "any conclusion…has a wider inductive basis when inferred from an experiment in which the quantities of other ingredients have been varied, than it would have from any amount of experimentation, in which these had been kept strictly constant" (p. 106). Fourthly, further subsidiary treatment factors can also be investigated. In summary, (i) knowledge of interactions, (ii) precision of treatment effects, (iii) generalisability of conclusions, and (iv) an ability to include subsidiary treatment factors were identified as the four benefits of conducting a factorial experiment.

### 3.1.1 Definitions of Treatment Effects

Yates[21,22] defined the treatment effects of interest in a factorial experiment, starting with a 2×2 factorial design. We label the two treatment factors $A$ and $B$. Following guidance from the MOST approach to clinical trials, we label the two levels of each factor $-1$ for low, absent or standard and $+1$ for high, present or enhanced.[18,19] We refer to the four treatments as ($-1$, $-1$) for $A_1$ and $B_1$, ($+1$, $-1$) for $A_2$ and $B_1$, ($-1$, $+1$) for $A_1$ and $B_2$, and ($+1$, $+1$) for $A_2$ and $B_2$. After Yates,[21,22] we denote the mean patient outcome for each treatment as $(1)$ for treatment ($-1$, $-1$), $a$ for treatment ($+1$, $-1$), $b$ for treatment ($-1$, $+1$) and finally $ab$ for treatment ($+1$, $+1$). Following Fisher,[17] Yates[21,22] defined two mean differences in outcome, which give information on the effect of factor $A$. Assuming the two levels of each treatment factor are present and absent, these are $\theta_{A1} = a - (1)$ for the effect of factor $A$ in the absence of $B$, and $\theta_{A2} = ab - b$ for the effect of $A$ in the presence of $B$. Yates[21] defined a "general" or "marginal" effect of factor $A$, which he labelled the "main effect" of $A$, as the average of these two "separate" or "conditional" effects, $\theta_{A1}$ and $\theta_{A2}$, as follows

$$\hat{A} = \tfrac{1}{2}\{(\mathrm{ab} - \mathrm{b}) + (\mathrm{a} - (1))\} = \tfrac{1}{2}\{\theta_{A1} + \theta_{A2}\}.$$

Likewise, Yates[21,22] defined two mean differences in outcome that give information on the effect of factor $B$, as $\theta_{B1} = \mathrm{b} - (1)$ for the effect of $B$ in the absence of $A$, and $\theta_{B2} = \mathrm{ab} - \mathrm{a}$ for the effect of $B$ in the presence of $A$. Yates[21,22] defined the main effect of factor $B$ as the average of these two separate effects, $\theta_{B1}$ and $\theta_{B2}$, as follows

$$\hat{B} = \tfrac{1}{2}\{(\mathrm{ab} - \mathrm{a}) + (\mathrm{b} - (1))\} = \tfrac{1}{2}\{\theta_{B1} + \theta_{B2}\}.$$

Finally, Yates[21,22] defined the $AB$ interaction as measuring the lack of additivity of these two main effects. He gave the $AB$ interaction as half the difference of the two separate effects of $A$ ($\theta_{A1}$ and $\theta_{A2}$), or identically, half the difference of the two separate effects of $B$ ($\theta_{B1}$ and $\theta_{B2}$),

$$\widehat{AB} = \tfrac{1}{2}\{(\mathrm{ab} - \mathrm{b}) - (\mathrm{a} - (1))\} = \tfrac{1}{2}\{(\mathrm{ab} - \mathrm{a}) - (\mathrm{b} - (1))\} = \tfrac{1}{2}\{\theta_{A2} - \theta_{A1}\} = \tfrac{1}{2}\{\theta_{B2} - \theta_{B1}\}.$$

For a summary, see Table 2 below.

[Insert Table 2 about here]

Yates[21] added the half to this definition of $\widehat{AB}$ for two reasons. Firstly, it follows that $\hat{A}$, $\hat{B}$ and $\widehat{AB}$ are estimated with equal precision. Secondly, it allows mean differences in outcome between treatments to be determined by “direct addition or subtraction” of the relevant main effect and interaction (p. 190). In a 2×2 factorial design, six possible pairwise differences in mean outcome between treatments (i.e., separate effects) are,

$$\begin{aligned}
\theta_{A1} &= \mathrm{a} - (1) = \hat{A} - \widehat{AB} \\
\theta_{A2} &= \mathrm{ab} - \mathrm{b} = \hat{A} + \widehat{AB} \\
\theta_{B1} &= \mathrm{b} - (1) = \hat{B} - \widehat{AB} \\
\theta_{B2} &= \mathrm{ab} - \mathrm{a} = \hat{B} + \widehat{AB} \\
\theta_{AB1} &= \mathrm{ab} - (1) = \hat{A} + \hat{B} \\
\theta_{AB2} &= \mathrm{a} - \mathrm{b} = \hat{A} - \hat{B}
\end{aligned}$$

Note that only three of the above effects are linearly independent. That is, given $\theta_{A1}$, $\theta_{A2}$ and $\theta_{AB1}$, $\theta_{B1}$, $\theta_{B2}$ and $\theta_{AB2}$ are fully determined.

Yates[21] noted that main effects are differences in the marginal totals, and the interaction the cross difference, in a two-way table of mean outcomes (p. 191). He observed that $\hat{A}$, $\hat{B}$ and $\widehat{AB}$ are all statistically independent, referring to them as orthogonal contrasts (p. 192). He[21] noted that "if there is no evidence of interaction, the mean [outcomes of] the two factors may be taken as the appropriate measures of the [outcomes of these factors, and thus] regarded as additive". He continued, if "interaction exists, then usually information will be required on the [outcomes for] each factor in the presence and absence of the other…[a factorial] experiment furnishes information on this point…with $\sqrt{2}$ times the standard error" (p. 193). Yates[21] concludes by claiming "[n]othing is lost…as even should the results be judged insufficiently precise, the information already obtained can be combined with that added later" (p. 193).

Moving onto the 2×2×2 factorial design as an example of a higher-order factorial design, we label the three treatment factors $A$, $B$ and $C$ and the two levels $-1$ and $+1$, and give the eight treatments, their associated labels and mean outcomes in Table 3, assuming a continuous outcome here as well.

[Insert Table 3 about here]

Here, Yates[21] defines the main effect of $A$ as the average of *four* separate effects that give information on the effect of factor $A$. These are $\theta_{A1} = \mathrm{a} - (1)$, the effect of $A$ in the absence of $B$ and $C$; $\theta_{A2} = \mathrm{ab} - \mathrm{b}$, the effect of $A$ in the presence of $B$ but absence of $C$; $\theta_{A3} = \mathrm{ac} - \mathrm{c}$, the effect of $A$ in the presence of $C$ but absence of $B$; and $\theta_{A4} = \mathrm{abc} - \mathrm{bc}$, the effect of $A$ in the presence of $B$ and $C$. This gives,

$$\hat{A} = \tfrac{1}{4}\{(\mathrm{abc} - \mathrm{bc}) + (\mathrm{ac} - \mathrm{c}) + (\mathrm{ab} - \mathrm{b}) + (\mathrm{a} - (1))\} = \tfrac{1}{4}\{\theta_{A1} + \theta_{A2} + \theta_{A3} + \theta_{A4}\}.$$

The main effects $\hat{B}$ and $\hat{C}$ are defined similarly. Yates[21] defines the first-order interaction of factors $A$ and $B$ as an average of two separate interactions in the presence and absence of $C$. These are $\theta_{AB1} = \tfrac{1}{2}\left((\mathrm{ab} - \mathrm{b}) - (\mathrm{a} - (1))\right)$ and $\theta_{AB2} = \tfrac{1}{2}\left((\mathrm{abc} - \mathrm{bc}) - (\mathrm{ac} - \mathrm{c})\right)$, giving

$$\widehat{AB} = \tfrac{1}{2}\left\{\tfrac{1}{2}\left((\mathrm{abc} - \mathrm{bc}) - (\mathrm{ac} - \mathrm{c})\right) + \tfrac{1}{2}\left((\mathrm{ab} - \mathrm{b}) - (\mathrm{a} - (1))\right)\right\} = \tfrac{1}{2}\{\theta_{AB1} + \theta_{AB2}\}.$$

The $\widehat{AC}$ and $\widehat{BC}$ interactions are defined similarly. Finally, Yates[21] defines a second-order interaction of factors $A$, $B$ and $C$ as half the difference between these two separate interactions $\theta_{AB1}$ and $\theta_{AB2}$, given by

$$\widehat{ABC} = \frac{1}{2}\left\{\frac{1}{2}\left((\mathrm{abc} - \mathrm{bc}) - (\mathrm{ac} - \mathrm{c})\right) - \frac{1}{2}\left((\mathrm{ab} - \mathrm{b}) - (\mathrm{a} - (1))\right)\right\}$$
$$= \tfrac{1}{2}\{\theta_{AB2} - \theta_{AB1}\} = \tfrac{1}{2}\{\theta_{AC2} - \theta_{AC1}\} = \tfrac{1}{2}\{\theta_{BC2} - \theta_{BC1}\}.$$

As before the main effects and interactions are all estimated with equal precision and separate effects are determined by adding or subtracting relevant main effects and interactions. So, for example, seven of a possible 28 separate effects are

$$\theta_{A1} = \mathrm{a} - (1) = \hat{A} - \widehat{AB} - \widehat{AC} + \widehat{ABC}$$
$$\theta_{B1} = \mathrm{b} - (1) = \hat{B} - \widehat{AB} - \widehat{BC} + \widehat{ABC}$$
$$\theta_{C1} = \mathrm{c} - (1) = \hat{C} - \widehat{AC} - \widehat{BC} + \widehat{ABC}$$
$$\theta_{AB1} = \mathrm{ab} - (1) = \hat{A} + \hat{B} - \widehat{AC} - \widehat{BC}$$
$$\theta_{AC1} = \mathrm{ac} - (1) = \hat{A} + \hat{C} - \widehat{AB} - \widehat{BC}$$
$$\theta_{BC1} = \mathrm{bc} - (1) = \hat{B} + \hat{C} - \widehat{AB} - \widehat{AC}$$
$$\theta_{ABC1} = \mathrm{abc} - (1) = \hat{A} + \hat{B} + \hat{C} + \widehat{ABC}$$

Twenty-one more separate effects are possible, with only seven of the 28 being linearly independent.

Yates[22] (p. 11) gives expressions for main effects and interactions from a 2×2×2 factorial experiment in terms of mean outcomes for each treatment, as found in Table 4. Note that there are always pluses in all rows for the grand mean. For the main effect of treatment factor $A$, because "$a$" is in the $a$, $ab, ac$ and $abc$ mean outcomes, a plus is in each of these rows for the $\hat{A}$ column, and minuses otherwise. The pluses and minuses are given using the same logic for the $\hat{B}$ and $\hat{C}$ columns. For the $AB$ interaction, the pluses and minuses are multiplied from the $\hat{A}$ and $\hat{B}$ columns to give the pluses and minuses in the rows for the $\widehat{AB}$ column. The same logic is used for the $\widehat{AC}$ and $\widehat{BC}$ columns. For the $ABC$ interaction, the pluses and minuses are multiplied from the $\widehat{AB}$ and $\hat{C}$ columns. This table can be used to derive all the separate effects as well. In a 2×2×2 factorial experiment, the 28 separate effects are ($\theta_{A1}$, $\theta_{A2}$, $\theta_{A3}$, $\theta_{A4}$, $\theta_{B1}$, $\theta_{B2}$, $\theta_{B3}$, $\theta_{B4}$, $\theta_{C1}$, $\theta_{C2}$, $\theta_{C3}$, $\theta_{C4}$, $\theta_{AB1}$, $\theta_{AB2}$, $\theta_{AB3}$, $\theta_{AB4}$, $\theta_{AC1}$, $\theta_{AC2}$, $\theta_{AC3}$, $\theta_{AC4}$ $\theta_{BC1}$, $\theta_{BC2}$, $\theta_{BC3}$, $\theta_{BC4}$, $\theta_{ABC1}$, $\theta_{ABC2}$, $\theta_{ABC3}$ and $\theta_{ABC4}$).

[Insert Table 4 about here]

### 3.1.2 Properties of Estimators

The properties of the estimators of these treatment effects have been written about, and their implications demonstrated, in relation to the MOST framework on clinical trials (in particular, see Chapter 6 in[19]). We set these out in mathematical terms below, giving the least squares estimators of the main effects and interactions, along with their variances. For simplicity, we assume an equal number of replicates for each treatment (i.e., a balanced design), so results can be derived for just one replicate. Least squares estimators for $r$ replicates are found by replacing the results from the single replicate with the mean outcomes from each treatment.

Least squares variances of the estimators are found for $r$ replicates by dividing the results for the single replicate by the number of replicates.

We use $\boldsymbol{X}$ and $\boldsymbol{y}$ to denote the model matrix (which corresponds to the pluses and minuses in the appropriate Table 4) and outcomes vector, respectively. The vector of estimators is given by $\hat{\beta}^{model} = (\boldsymbol{X}'\boldsymbol{X})^{-1}\boldsymbol{X}'\boldsymbol{y}$. The variance-covariance matrix $Cov(\hat{\beta}^{model})$ for those estimators is given by $\sigma^2(\boldsymbol{X}'\boldsymbol{X})^{-1}$. For illustration, we define two of the possible models. In the *maximal* model, also referred to as the full model, all main effects and interactions are included. In the *additive* model, only the main effects are included.

For a single replicate of the 2×2 factorial design, the additive model $\boldsymbol{X}$ and $\boldsymbol{y}$ are,

$$\boldsymbol{X} = \begin{bmatrix} 1 & -1 & -1 \\ 1 & 1 & -1 \\ 1 & -1 & 1 \\ 1 & 1 & 1 \end{bmatrix} \quad \text{and} \quad \boldsymbol{y} = \begin{bmatrix} y_1 \\ y_2 \\ y_3 \\ y_4 \end{bmatrix}$$

$$\text{so} \quad (\boldsymbol{X}'\boldsymbol{X})^{-1} = \begin{bmatrix} \frac{1}{4} & 0 & 0 \\ 0 & \frac{1}{4} & 0 \\ 0 & 0 & \frac{1}{4} \end{bmatrix} \quad \text{and} \quad \boldsymbol{X}'\boldsymbol{y} = \begin{bmatrix} y_1 + y_2 + y_3 + y_4 \\ (y_2 + y_4) - (y_1 + y_3) \\ (y_3 + y_4) - (y_1 + y_2) \end{bmatrix} \text{ and so}$$

$$\hat{\beta}^{additive} = \begin{bmatrix} \hat{\mu} \\ \hat{A}/2 \\ \hat{B}/2 \end{bmatrix} = \begin{bmatrix} \frac{1}{4}(y_1 + y_2 + y_3 + y_4) \\ \frac{1}{4}(y_2 + y_4) - \frac{1}{4}(y_1 + y_3) \\ \frac{1}{4}(y_3 + y_4) - \frac{1}{4}(y_1 + y_2) \end{bmatrix} \quad \text{with } Cov(\hat{\beta}^{additive}) = \begin{bmatrix} \frac{\sigma^2}{4} & 0 & 0 \\ 0 & \frac{\sigma^2}{4} & 0 \\ 0 & 0 & \frac{\sigma^2}{4} \end{bmatrix}.$$

As all the off-diagonal elements of $Cov(\hat{\beta}^{additive})$ are zero, the estimators are orthogonal (or uncorrelated).

For the maximal model, $\boldsymbol{X}$ and $\boldsymbol{y}$ are,

$$\boldsymbol{X} = \begin{bmatrix} 1 & -1 & -1 & 1 \\ 1 & 1 & -1 & -1 \\ 1 & -1 & 1 & -1 \\ 1 & 1 & 1 & 1 \end{bmatrix} \quad \text{and} \quad \boldsymbol{y} = \begin{bmatrix} y_1 \\ y_2 \\ y_3 \\ y_4 \end{bmatrix}$$

$$\text{so} \quad \hat{\beta}^{max} = \begin{bmatrix} \hat{\mu} \\ \hat{A}/2 \\ \hat{B}/2 \\ \widehat{AB}/2 \end{bmatrix} = \begin{bmatrix} \frac{1}{4}(y_1 + y_2 + y_3 + y_4) \\ \frac{1}{4}(y_2 + y_4) - \frac{1}{4}(y_1 + y_3) \\ \frac{1}{4}(y_3 + y_4) - \frac{1}{4}(y_1 + y_2) \\ \frac{1}{4}(y_1 + y_4) - \frac{1}{4}(y_2 + y_3) \end{bmatrix} \quad \text{with } Cov(\hat{\beta}^{max}) = \begin{bmatrix} \frac{\sigma^2}{4} & 0 & 0 & 0 \\ 0 & \frac{\sigma^2}{4} & 0 & 0 \\ 0 & 0 & \frac{\sigma^2}{4} & 0 \\ 0 & 0 & 0 & \frac{\sigma^2}{4} \end{bmatrix}.$$

The estimators of the main effects ($\hat{A}$ and $\hat{B}$) and their variances are unchanged by adding the interaction ($\widehat{AB}$) into the model. The standard errors of $\hat{A}$, $\hat{B}$ and $\widehat{AB}$ are all equal as expected.

We turn now to a 2×2×2 factorial design. For a single replicate, $\boldsymbol{X}$ and $\boldsymbol{y}$ are,

$$\boldsymbol{X} = \begin{bmatrix} 1 & -1 & -1 & 1 & -1 & 1 & 1 & -1 \\ 1 & 1 & -1 & -1 & -1 & -1 & 1 & 1 \\ 1 & -1 & 1 & -1 & -1 & 1 & -1 & 1 \\ 1 & 1 & 1 & 1 & -1 & -1 & -1 & -1 \\ 1 & -1 & -1 & 1 & 1 & -1 & -1 & 1 \\ 1 & 1 & -1 & -1 & 1 & 1 & -1 & -1 \\ 1 & -1 & 1 & -1 & 1 & -1 & 1 & -1 \\ 1 & 1 & 1 & 1 & 1 & 1 & 1 & 1 \end{bmatrix} \quad \text{and} \quad \boldsymbol{y} = \begin{bmatrix} y_1 \\ y_2 \\ y_3 \\ y_4 \\ y_5 \\ y_6 \\ y_7 \\ y_8 \end{bmatrix}.$$

$$\text{so} \quad \hat{\beta}^{max} = \begin{bmatrix} \hat{\mu} \\ \hat{A}/2 \\ \hat{B}/2 \\ \widehat{AB}/2 \\ \hat{C}/2 \\ \widehat{AC}/2 \\ \widehat{BC}/2 \\ \widehat{ABC}/2 \end{bmatrix} = \begin{bmatrix} \frac{1}{8}(y_1+y_2+y_3+y_4+y_5+y_6+y_7+y_8) \\ \frac{1}{8}(y_2+y_4+y_6+y_8) - \frac{1}{8}(y_1+y_3+y_5+y_7) \\ \frac{1}{8}(y_3+y_4+y_7+y_8) - \frac{1}{8}(y_1+y_2+y_5+y_6) \\ \frac{1}{8}(y_1+y_4+y_5+y_8) - \frac{1}{8}(y_2+y_3+y_6+y_7) \\ \frac{1}{8}(y_5+y_6+y_7+y_8) - \frac{1}{8}(y_1+y_2+y_3+y_4) \\ \frac{1}{8}(y_1+y_3+y_6+y_8) - \frac{1}{8}(y_2+y_4+y_5+y_7) \\ \frac{1}{8}(y_1+y_2+y_7+y_8) - \frac{1}{8}(y_3+y_4+y_5+y_6) \\ \frac{1}{8}(y_2+y_3+y_5+y_8) - \frac{1}{8}(y_1+y_4+y_6+y_7) \end{bmatrix}$$

$$\text{and} \quad Cov(\hat{\beta}^{max}) = \begin{bmatrix} \frac{\sigma^2}{8} & 0 & 0 & 0 & 0 & 0 & 0 & 0 \\ 0 & \frac{\sigma^2}{8} & 0 & 0 & 0 & 0 & 0 & 0 \\ 0 & 0 & \frac{\sigma^2}{8} & 0 & 0 & 0 & 0 & 0 \\ 0 & 0 & 0 & \frac{\sigma^2}{8} & 0 & 0 & 0 & 0 \\ 0 & 0 & 0 & 0 & \frac{\sigma^2}{8} & 0 & 0 & 0 \\ 0 & 0 & 0 & 0 & 0 & \frac{\sigma^2}{8} & 0 & 0 \\ 0 & 0 & 0 & 0 & 0 & 0 & \frac{\sigma^2}{8} & 0 \\ 0 & 0 & 0 & 0 & 0 & 0 & 0 & \frac{\sigma^2}{8} \end{bmatrix}.$$

The estimators and their variances are again unchanged by adding interactions into the model; the standard errors are again all equal. As the ratio of the estimators and their standard errors is the same in a 2×2 and 2×2×2 factorial design, main effects and interactions are estimated with equal precision, if the sample size remains the same.

### 3.1.3 Approaches to Decision Making

More than one approach has been recommended for making decisions from a Design of Experiments perspective. Collins *et al*[41] suggested a component screening approach for deciding which components to include in an optimised intervention, which is close to the tradition of the "Wisconsin school" including.[5,28] First, the main effects are examined to determine whether there is evidence of a general effect of each component in the desired direction. If an effect is positive, each component is tentatively screened in, otherwise it is tentatively screened out. Second, tentative decisions are reconsidered in light of interaction effects, starting with lower-order interactions. If the effect of a screened in component diminishes sufficiently when combined with another component, it may be moved to the screened-out set. Likewise, if the effect of a screened-out component increases sufficiently when it is combined with another component, it may be moved to the screened in set. This approach respects a hierarchical ordering principle giving more emphasis to lower-order effects. Collins *et al*[41] relaxed the heredity principle stating interactions are only of interest if all the components have sufficiently large main effects to a requirement for at least one of the main effects being sufficiently large. What is sufficiently large is based on *a priori* thresholds,[41] not necessarily hypothesis testing. A third and final step is to take the constraints of affordability, scalability and efficiency into account, explicitly balancing effectiveness against these key constraints.[18,32]

In an alternative approach, Bailey[23] suggests that the maximal model is fitted. If and only if there is interest in fitting different models for the fixed effects (i.e., treatment effects), she recommends using hypothesis tests to select the smallest or most parsimonious model the data support and then estimating the parameters from this model (p. 84). Three principles are proposed to choose the models considered for the fixed effects.[23] Firstly, the 'intersection' principle states that if two models are to be considered, a third model representing their intersection should also be considered. So, if two models are considered, one with the grand mean and $\hat{A}$ and a second with the grand mean and $\hat{B}$, then a model with just the grand mean should also be considered. Secondly, the 'sum' principle states that where two models are considered, a third model that includes all fixed effects across both models should also be considered. This would be one with the grand mean, $\hat{A}$ and $\hat{B}$. Finally, the 'orthogonality' principle states that when two reduced models of a common maximal model are considered, the test for removing fixed effects from a reduced model should be the same as the test for removing them from the maximal model, interchanged for each set of fixed effects removed. Here, if the maximal model includes the grand mean, $\hat{A}$ and $\hat{B}$, then the test for removing $\hat{A}$ from this maximal model should be the same as the test for removing $\hat{A}$ from a model with just the grand mean and $\hat{A}$. This ensures all possible model selection procedures ultimately give the same reduced model.

Once a collection of models of interest is selected, Bailey[23] recommends starting with the maximal model (i.e., backwards selection) and stopping with the smallest model that is supported by the data. As a result, higher-order interactions, then lower-order interactions, and then main effects are removed. The principle of 'marginality' states that once a treatment effect is classified as important, all the lower-order effects related to it must be kept in the model, regardless of their individual importance. The final model is then used to predict the optimal treatment.

The component screening approach is no longer recommended within MOST.[32] Instead, a new approach referred to as decision analysis for intervention value efficiency (or DAIVE) is advocated, which builds on Bayesian decision theory and multi-criteria decision analysis.[42] The advantage of this alternative approach is that it selects an optimal intervention based on the posterior expected values of all treatments under consideration so is less laborious. It also accommodates more than one outcome, allowing the contribution of each outcome to the final decision to be weighted.[42]

## 3.2 *k*-in-1 Factorial Clinical Trials

Neuhauser[43] attributed use of factorial designs in clinical trials being rare until the 1980s to Bradford Hill's decision to promote use of simple experimental methods in medicine. Early in the twentieth century, clinical trials were carried out by clinicians with limited statistical input.[43] Bradford Hill felt clinicians would "flinch from anything but the simplest experimental methods" (p. 109),[43] helping to establish two-arm parallel-group designs as the standard. In a departure from this, Peto[7] argued factorial designs are one of the few substantial advances in clinical trial design which "can be implemented with little or no extra difficulty or cost" (p. 34). He argued that, if there are "two largely unrelated questions…concerning the treatment of the same disease" (p. 34), it would be more efficient to use a 2×2 factorial design than to answer one question at a time or to randomise three ways.[7] He saw no reason why 2×2×2 or $2^4$ factorial designs should "not be considered to answer 3 or even 4 questions simultaneously and efficiently" (p. 35).[7] In this way, Peto[7] built on Fisher's[17] argument that factorial experiments enable subsidiary treatment factors to be added.

Brittain and Wittes[44] later described four rationales for conducting a factorial clinical trial. These are (i) to modify an already multipronged approach to therapy, (ii) to test a new therapy if another unproven therapy is already commonly used, (iii) if different outcomes are used for different treatment factors, and (iv) to test several new or previously untested therapies with a single outcome.[44] The fourth rationale was singled out as relevant to *k*-in-1 factorial trials, with the first three making a factorial clinical trial clearly desirable. Brittain and Wittes[44] argue that if two new therapies are to be included with a single outcome, a three-arm parallel-group trial design is an alternative design, and a 2-in-1 factorial design may (or may not) be more efficient than this. Brittain and Wittes[44] state the objective of a *k*-in-1 factorial trial as the estimation of the "effect of each treatment when administered alone" (p. 162). That is, the effects of just factor A and B in the absence of the other. Estimation of the effects of factor A and B in the presence of the other, averaged across the presence and absence of the other, or in combination with each other are not usually primary objectives of a *k*-in-1 factorial trial.

By 2002, Green *et al*,[45] noting the importance of considering the goal of a factorial trial, argued "[o]ccasionally…two primary questions…may be considered independent questions. More often…the goal at the end…is to recommend for use one of the four treatment arms" (p. 3425). Hence, Green *et al*[45] conflated the purpose of a 2-in-1 factorial trial, with that of a factorial evaluation trial. To clarify, the reason for including all four treatments in a 2-in-1 factorial trial is the efficiency gained by the treatment contrasts possible relative to those that are possible in two separate two-arm parallel-group trials or a single three-arm parallel-group trial. The estimands of interest remain the same and so the sample size is based on the effects of factor A or B in the absence of the other (estimated from the main effect assuming the absence of interaction). The reason for including all four treatments in a factorial evaluation trial is that the treatments then have what is referred to as a factorial treatment structure (see[23] for example), which enables estimation of two main effects and an interaction. As such, the estimands of interest change and the sample size is now based on detecting the two main effects and the interaction. We suggest that the term "factorial design" should not be used alone, as the same four treatments can be randomised identically whether the clinical trial is referred to as a 4-arm parallel-group or a 2×2 factorial experiment. Instead, we suggest that the research questions of primary interest, and the resulting estimands, are clearly specified and the design is referred to as a 4-arm parallel-group, a 2-in-1 factorial, a 2×2 factorial screening, a 2×2 factorial optimisation, or a 2×2 factorial evaluation design.

### 3.2.1 Definitions of Treatment Effects

In 1989, Brittain and Wittes[44] formally set out the treatment effects of interest in factorial clinical trials conducted for different purposes. Where the purpose is to modify an already multipronged approach to therapy, they argued that the main effects provide an "appropriate" or "useful measure of the expected size of the treatment effect" when the range of treatments in the trial reflects the range found in medical practice (pp. 161-162). They state that the usual parameterisation of the analysis (by which they mean that given in Section 3.1) gives the main effects.[44] However, in a 2-in-1 factorial trial, Brittain and Wittes believed an alternative parameterisation to be "more natural".[44] Specifically, they suggested the use of an "additive model" that "compares each treatment to placebo" (p. 164). In this way, they proposed that a model is fitted, including only the effects of $A$ and $B$, but with different treatment contrasts. They clarify that the parameters are then interpreted as "the control mean and the effects of treatments A and B when administered without the other" (p. 164). In this way, two treatment contrasts are proposed but these are not the orthogonal contrasts proposed by Yates.[21,22] The purpose of a 2-in-1 factorial trial is explicitly stated to be estimation of two separate effects of factors $A$ and $B$ (i.e., $\theta_{A1} = \mathrm{a} - (1)$ and $\theta_{B1} = \mathrm{b} - (1)$), but using all the available data (and fitting marginal effects, instead of using half the data and fitting the separate effects directly).

Recently, Kahan *et al*[46] formally defined the estimands they considered to be of potential interest using the ICH-E9(R1) addendum on estimands.[47] They argued that any subset of the following estimands may be of interest for the effect of $A$: i) the effect of treatment A ($A_2$, $B_1$) versus control ($A_1, B_1$) in the absence of treatment B ($\theta_{A1} = \mathrm{a} - (1)$) (i.e., if no one receives treatment B), ii) the effect of treatment A ($A_2, B_2$) versus control ($A_1, B_2$) in the presence of treatment B ($\theta_{A2} = \mathrm{ab} - \mathrm{b}$) (i.e., if everyone receives treatment B), iii) the effect of treatment A ($A_2, \breve{B}$) versus control ($A_1, \breve{B}$) if treatment B is given according to usual practice (not previously defined) (i.e., if everyone received treatment B as they would under usual practice), and iv) the effect of treatment A in combination with treatment B ($A_2, B_2$) vs control ($A_1, B_1$) ($\theta_{AB1} = ab - (1)$) (i.e., the effect of both A and B against neither treatment A nor B) (p. 3). Kahan *et al*[46] note "the estimands defined here could equally be applied to the setting where only the effect of treatment A is of interest (e.g., a two-arm [parallel-group] trial where treatment B is a concomitant treatment)" (p. 2). By omission, the main effects $\hat{A}$ and $\hat{B}$ and interaction $\widehat{AB}$ proposed by Fisher and Yates[17,20,21,22] are not believed to be estimands of interest in a 2-in-1 factorial trial despite the focus on estimating marginal effects.

### 3.2.2 Properties of Estimators

The estimands proposed for 2-in-1 factorial trials[44,46] are directly estimated from specific pairwise comparisons between treatments (separate effects). Importantly though, the effects of $A$ and $B$ when administered alone ($A_2, B_1$ and $A_1, B_2$) compared to control (i.e., $A_1, B_1$) can be directly estimated more efficiently using the additive model, including the marginal effects of $A$ and $B$ with dummy-coded treatment effects, if the interaction between the treatment factors is assumed to be zero. That is, if the separate effects of $A$ ($\theta_{A1}$ and $\theta_{A2}$) are equal, all the data can be used, and instead of the main effect $\hat{A}$ being equal to $(\theta_{A1} + \theta_{A2})/2$, it now equals $\hat{A} = \theta_{A1} = \theta_{A2}$. Equally, if the separate effects of $B$ ($\theta_{B1}$ and $\theta_{B2}$) are equal, all the data can be used, and instead of the main effect $\hat{B}$ being equal to $(\theta_{B1} + \theta_{B2})/2$, it now equals $\hat{B} =$

$\theta_{B1} = \theta_{B2}$.

Two models can again be defined. In the additive model suggested by Brittain and Wittes,[44] an intercept and marginal effects of $A$ and $B$ are included. These marginal effects ($\theta_{A1}$ and $\theta_{B1}$) (under dummy coding) are twice the size of the main effects obtained directly from the regression (under effect coding) because there is a one-unit difference between 0 and 1 and a two-unit difference between $-1$ and $+1$. This additive model is often referred to as the "factorial analysis" in the clinical trials literature (e.g.,[46]). In one parameterisation of the maximal model, the effect of $A$ and $B$ combined ($\theta_{AB1}$) is also included. This is referred to as a "multi-arm analysis" in.[46]

In a 2-in-1 factorial trial, following guidance from the MOST, we label the two levels of each treatment factor zero for low, absent or standard and one for high, present or enhanced.[18,19] We then refer to the four treatments as (0, 0) for $A_1$ and $B_1$, (1, 0) for $A_2$ and $B_1$, (0, 1) for $A_1$ and $B_2$, and (1, 1) for $A_2$ and $B_2$. This difference in labelling is consistent with using dummy (0, 1) coding as opposed to effect $(-1, +1)$ coding of treatment variables in analysing a *k*-in-1 factorial trial.[18,19] This enables the regression parameters to directly provide the treatment contrasts of interest. Dummy coding gives marginal and separate effects and effect coding gives main effects and interactions. In some Design of Experiments textbooks (e.g.,[23]) and in agricultural experiments, it is standard to dummy code treatment variables and then set up the main effects and interactions afterwards as functions of parameters. This is less standard in clinical trials, hence the emphasis in MOST[18,19] on coding treatment variables. While the labels given to the levels of the treatment factors are arbitrary, and the treatment contrasts specified in the model matrix important, the practical guidance from[18,19] is helpful.

Piantadosi[48] provides least squares estimators for the effects of $A$ and $B$ from Brittain and Wittes'[44] additive model, along with their variances. The model matrix and outcome vector for a single replicate are specified as follows, with the choice of model parameterisation and constraints made explicit,

$$\boldsymbol{X} = \begin{bmatrix} 1 & 0 & 0 \\ 1 & 1 & 0 \\ 1 & 0 & 1 \\ 1 & 1 & 1 \end{bmatrix} \text{ and } \boldsymbol{y} = \begin{bmatrix} y_1 \\ y_2 \\ y_3 \\ y_4 \end{bmatrix}$$

$$\text{so } (\boldsymbol{X}'\boldsymbol{X})^{-1} = \begin{bmatrix} \frac{3}{4} & -\frac{1}{2} & -\frac{1}{2} \\ -\frac{1}{2} & 1 & 0 \\ -\frac{1}{2} & 0 & 1 \end{bmatrix} \text{ and } \boldsymbol{X}'\boldsymbol{y} = \begin{bmatrix} y_1 + y_2 + y_3 + y_4 \\ (y_2 + y_4) \\ (y_3 + y_4) \end{bmatrix} \text{ so}$$

$$\tilde{\beta}^{additive} = \begin{bmatrix} \tilde{\mu} \\ \tilde{A} \\ \tilde{B} \end{bmatrix} = \begin{bmatrix} \frac{3}{4}y_1 + \frac{1}{4}y_2 + \frac{1}{4}y_3 - \frac{1}{4}y_4 \\ \frac{1}{2}(y_2 + y_4) - \frac{1}{2}(y_1 + y_3) \\ \frac{1}{2}(y_3 + y_4) - \frac{1}{2}(y_1 + y_2) \end{bmatrix} \text{ and } Cov(\tilde{\beta}^{additive}) = \begin{bmatrix} \frac{3\sigma^2}{4} & -\frac{\sigma^2}{2} & -\frac{\sigma^2}{2} \\ -\frac{\sigma^2}{2} & \sigma^2 & 0 \\ -\frac{\sigma^2}{2} & 0 & \sigma^2 \end{bmatrix}.$$

Note that in this parameterisation of the additive model, $\tilde{\mu} \neq \hat{\mu}$, $\tilde{A} = \hat{A}$ and $\tilde{B} = \hat{B}$, directly estimating the main effects defined by Yates.[21] The off-diagonal elements of $Cov(\tilde{\beta}^{additive})$ are zero or $-\sigma^2/2$, which indicates $\tilde{A}$ and $\tilde{B}$ are orthogonal contrasts, but $\tilde{\mu}$ is now correlated with $\tilde{A}$ and $\tilde{B}$. As $Var(\tilde{A}) = 2^2 Var(\hat{A}/2)$ and $Var(\tilde{B}) = 2^2 Var(\hat{B}/2)$, the precision of $\tilde{A}$ and $\tilde{B}$ is proportional to the precision of $\hat{A}/2$ and $\hat{B}/2$, so the hypothesis tests are the same.

The corresponding maximal model matrix and outcomes vector for a single replicate are specified as follows, the choice of model parameterisation and constraints again being made explicit,

$$\boldsymbol{X} = \begin{bmatrix} 1 & 0 & 0 & 0 \\ 1 & 1 & 0 & 0 \\ 1 & 0 & 1 & 0 \\ 1 & 1 & 1 & 1 \end{bmatrix} \text{ and } \boldsymbol{y} = \begin{bmatrix} y_1 \\ y_2 \\ y_3 \\ y_4 \end{bmatrix}, \text{ so that}$$

$$\tilde{\beta}^{max} = \begin{bmatrix} \tilde{\mu} \\ \tilde{A} \\ \tilde{B} \\ \widetilde{AB} \end{bmatrix} = \begin{bmatrix} y_1 \\ y_2 - y_1 \\ y_3 - y_1 \\ (y_1 + y_4) - (y_2 + y_3) \end{bmatrix}, \text{ and}$$

$$Cov(\tilde{\beta}^{max}) = \begin{bmatrix} \sigma^2 & -\sigma^2 & -\sigma^2 & \sigma^2 \\ -\sigma^2 & 2\sigma^2 & \sigma^2 & -2\sigma^2 \\ -\sigma^2 & \sigma^2 & 2\sigma^2 & -2\sigma^2 \\ \sigma^2 & -2\sigma^2 & -2\sigma^2 & 4\sigma^2 \end{bmatrix}.$$

Note that in this parameterisation of the maximal model, $\tilde{\mu} = (1)$, $\tilde{A} = \theta_{A1} = a - (1)$, $\tilde{B} = \theta_{B1} = b - (1)$ and $\widetilde{AB} = (1) + ab - a - b = 2\widehat{AB} \neq \theta_{AB}$. As such, $\tilde{A}$ and $\tilde{B}$ are estimated using only half the data, while $\widetilde{AB}$ is estimated using all the data. Note also that the standard errors of $\tilde{A}$ and $\tilde{B}$ from $Cov(\tilde{\beta}^{max})$ are $\sqrt{2}$ times those from $Cov(\tilde{\beta}^{additive})$. The variance of $\widetilde{AB}$ ($4\sigma^2$) is twice that of the variances of $\tilde{A}$ and $\tilde{B}$ ($2\sigma^2$), and $Var(\widetilde{AB}) = 4^2 Var(\widehat{AB}/2)$. Now, all the estimators are correlated with each other. The estimators and their variances are all changed by adding $\widetilde{AB}$ into the model.

Turning now to a 3-in-1 factorial trial, the additive model matrix and outcomes vector for a single replicate are specified as follows, again making the choice of model parameterisation and constraints explicit,

$$\boldsymbol{X} = \begin{bmatrix} 1 & 0 & 0 & 0 \\ 1 & 1 & 0 & 0 \\ 1 & 0 & 1 & 0 \\ 1 & 1 & 1 & 0 \\ 1 & 0 & 0 & 1 \\ 1 & 1 & 0 & 1 \\ 1 & 0 & 1 & 1 \\ 1 & 1 & 1 & 1 \end{bmatrix} \text{ and } \boldsymbol{y} = \begin{bmatrix} y_1 \\ y_2 \\ y_3 \\ y_4 \\ y_5 \\ y_6 \\ y_7 \\ y_8 \end{bmatrix}$$

so $$\tilde{\beta}^{additive} = \begin{bmatrix} \tilde{\mu} \\ \tilde{A} \\ \tilde{B} \\ \tilde{C} \end{bmatrix} = \begin{bmatrix} \frac{1}{2}y_1 + \frac{1}{4}y_2 + \frac{1}{4}y_3 + \frac{1}{4}y_5 - \frac{1}{4}y_8 \\ \frac{1}{4}(y_2 + y_4 + y_6 + y_8) - \frac{1}{4}(y_1 + y_3 + y_5 + y_7) \\ \frac{1}{4}(y_3 + y_4 + y_7 + y_8) - \frac{1}{4}(y_1 + y_2 + y_5 + y_6) \\ \frac{1}{4}(y_5 + y_6 + y_7 + y_8) - \frac{1}{4}(y_1 + y_2 + y_3 + y_4) \end{bmatrix}$$ and

$$Cov(\tilde{\beta}^{additive}) = \begin{bmatrix} \frac{\sigma^2}{2} & -\frac{\sigma^2}{4} & -\frac{\sigma^2}{4} & -\frac{\sigma^2}{4} \\ -\frac{\sigma^2}{4} & \frac{\sigma^2}{2} & 0 & 0 \\ -\frac{\sigma^2}{4} & 0 & \frac{\sigma^2}{2} & 0 \\ -\frac{\sigma^2}{4} & 0 & 0 & \frac{\sigma^2}{2} \end{bmatrix}.$$

Note that in this parameterisation of the additive model, $\tilde{\mu} \neq \hat{\mu}$, $\tilde{A} = \hat{A}$, $\tilde{B} = \hat{B}$ and $\tilde{C} = \hat{C}$, again directly estimating the main effects defined by Yates.[21] The diagonal elements of $Cov(\tilde{\beta}^{additive})$ and $Cov(\hat{\beta}^{additive})$ are proportional. It follows that the precisions of $\tilde{A}$, $\tilde{B}$ and $\tilde{C}$ are proportional to $\hat{A}/2$, $\hat{B}/2$ and $\hat{C}/2$, and the hypothesis tests are the same. Off-diagonal elements of $Cov(\tilde{\beta}^{additive})$ are zero or $-\sigma^2/4$, indicating that $\tilde{A}$, $\tilde{B}$ and $\tilde{C}$ are orthogonal contrasts, but $\tilde{\mu}$ is now correlated with $\tilde{A}$, $\tilde{B}$ and $\tilde{C}$.

The maximal model matrix and outcomes vector for a single replicate are specified below, with the choice of model parameterisation and constraints made explicit,

$$\boldsymbol{X} = \begin{bmatrix} 1 & 0 & 0 & 0 & 0 & 0 & 0 & 0 \\ 1 & 1 & 0 & 0 & 0 & 0 & 0 & 0 \\ 1 & 0 & 1 & 0 & 0 & 0 & 0 & 0 \\ 1 & 1 & 1 & 1 & 0 & 0 & 0 & 0 \\ 1 & 0 & 0 & 0 & 1 & 0 & 0 & 0 \\ 1 & 1 & 0 & 0 & 1 & 1 & 0 & 0 \\ 1 & 0 & 1 & 0 & 1 & 0 & 1 & 0 \\ 1 & 1 & 1 & 1 & 1 & 1 & 1 & 1 \end{bmatrix} \text{ and } \boldsymbol{y} = \begin{bmatrix} y_1 \\ y_2 \\ y_3 \\ y_4 \\ y_5 \\ y_6 \\ y_7 \\ y_8 \end{bmatrix}, \text{ so that}$$

$$\tilde{\beta}^{max} = \begin{bmatrix} \tilde{\mu} \\ \tilde{A} \\ \tilde{B} \\ \widetilde{AB} \\ \tilde{C} \\ \widetilde{AC} \\ \widetilde{BC} \\ \widetilde{ABC} \end{bmatrix} = \begin{bmatrix} y_1 \\ y_2 - y_1 \\ y_3 - y_1 \\ (y_1 + y_4) - (y_2 + y_3) \\ y_5 - y_1 \\ (y_1 + y_6) - (y_2 + y_5) \\ (y_1 + y_7) - (y_3 + y_5) \\ (y_2 + y_3 + y_5 + y_8) - (y_1 + y_4 + y_6 + y_7) \end{bmatrix}$$

and $$Var(\tilde{\beta}^{max}) = \begin{bmatrix} \sigma^2 & -\sigma^2 & -\sigma^2 & \sigma^2 & -\sigma^2 & \sigma^2 & \sigma^2 & -\sigma^2 \\ -\sigma^2 & 2\sigma^2 & \sigma^2 & -2\sigma^2 & \sigma^2 & -2\sigma^2 & -\sigma^2 & 2\sigma^2 \\ -\sigma^2 & \sigma^2 & 2\sigma^2 & -2\sigma^2 & \sigma^2 & -\sigma^2 & -2\sigma^2 & 2\sigma^2 \\ \sigma^2 & -2\sigma^2 & -2\sigma^2 & 4\sigma^2 & -\sigma^2 & -2\sigma^2 & -2\sigma^2 & -4\sigma^2 \\ -\sigma^2 & \sigma^2 & \sigma^2 & -\sigma^2 & 2\sigma^2 & -2\sigma^2 & -2\sigma^2 & 2\sigma^2 \\ \sigma^2 & -2\sigma^2 & -\sigma^2 & -2\sigma^2 & -2\sigma^2 & 4\sigma^2 & 2\sigma^2 & -4\sigma^2 \\ \sigma^2 & -\sigma^2 & -2\sigma^2 & -2\sigma^2 & -2\sigma^2 & 2\sigma^2 & 4\sigma^2 & -4\sigma^2 \\ -\sigma^2 & 2\sigma^2 & 2\sigma^2 & -4\sigma^2 & 2\sigma^2 & -4\sigma^2 & -4\sigma^2 & 8\sigma^2 \end{bmatrix}$$

Note that $\tilde{\mu} = (1)$, $\tilde{A} = \theta_{A1} = a - (1)$, $\tilde{B} = \theta_{B1} = b - (1)$, $\widetilde{AB} = (1) + ab - a - b \neq \theta_{AB1}$, $\tilde{C} = \theta_{C1} = c - (1)$, $\widetilde{AC} = (1) + ac - a - c \neq \theta_{AC1}$, $\widetilde{BC} = (1) + bc - b - c \neq \theta_{BC1}$ and $\widetilde{ABC} = abc + c + b + a - bc - ac - ab - (1) = 4\widehat{ABC} \neq \theta_{ABC}$. It is clear from this that $\tilde{A}$, $\tilde{B}$ and $\tilde{C}$ use only a quarter of the data and thus clearly differ from $\hat{A}$, $\hat{B}$ and $\hat{C}$. It is also clear that $\widetilde{AB}$, $\widetilde{AC}$ and $\widetilde{BC}$ use only half the data and thus clearly differ from $\widehat{AB}$, $\widehat{AC}$ and $\widehat{BC}$. Note also, that in the maximal model, the variance of $\tilde{A}, \tilde{B}$ and $\tilde{C}$ from $Cov(\tilde{\beta}^{max})$ are now four times those from $Cov(\tilde{\beta}^{additive})$. The variance of $\widetilde{AB}$, $\widetilde{AC}$ and $\widetilde{BC}$ is twice, and the variance of $\widetilde{ABC}$ four times the size of the variance of $\tilde{A}, \tilde{B}$ and $\tilde{C}$. This means that the precision with which separate interactions are estimated is now lower than the precision with which the separate effects are estimated. Importantly, $Var(\widetilde{ABC}) = 4^3 Var(\widehat{ABC}/2)$. All the estimators are now correlated, so all the estimators and their variances change by adding interactions into the model. In this way, setting out to fit the dummy-coded additive model, but then fitting the associated maximal model because the assumption of no interactions is found to be implausible, makes it unlikely there will be sufficient precision to draw clear conclusions.

### 3.2.3 Approaches to Decision Making

The approach recommended for making decisions in *k*-in-1 factorial trials is specific to their purpose and has evolved over time. Based on their simulation study, Green *et al*[45] argued that the analysis approach should depend on whether interactions are anticipated. Where large interactions are not anticipated, they recommended a two-stage testing procedure in which a preliminary test for interactions is conducted, and then the additive model recommended by Brittain and Wittes'[44] is fitted. When large interactions are anticipated, they recommended directly estimating the separate effects, powering the trial to detect plausible interactions if the interactions are of interest. McAlister *et al*,[6] reinforcing,[45] recommended that, "in the presence of interactions, analysis must be carried out "inside the table" [by which they mean directly estimating the separate effects]" (p. 2546). Unaware of MOST,[31] Montgomery *et al*[49] contrasted an "inside the table" and an "at the margins" analysis, implicitly assuming all trialists, including those conducting factorial trials of complex interventions, are interested only in estimating separate effects. Based on the potential for bias, Kahan[50] recommended that the dummy-coded maximal model should be fitted in addition to the additive model[44], as a sensitivity analysis. Korn and Freidlin[51] recommend that this dummy-coded maximal model is fitted as the primary analysis, protecting for multiple testing, whenever interactions are plausible. This guidance is sensible only in the context of *k*-in-1 factorial trials.

## 4. THE DEBATE ABOUT INTERACTIONS

There has been much debate on the role interactions play in factorial trials. In 1935, Neyman[52] argued that experimenters want to estimate what he referred to as "simple effects" (these are synonymous with separate effects) and "simple interactions" (synonymous with separate interactions). In a 2×2 factorial trial, these are $\tilde{A}$, $\tilde{B}$ and $\widetilde{AB}$. He[52] went on to show that the power to detect plausible simple effects is low, so a factorial experiment is "rather a poor experiment" (p. 237). Neyman states the "position with regard to interactions is still worse", with two-factor interactions being "practically undetectable" (p. 237).[52] Neyman concluded that factorial experiments work "very well if the interactions are non-existent. However, in cases where they do exist…the method may give unsatisfactory results." (p. 238).[52] He illustrates his argument using a theoretical case where an interaction exists but is not significant and the simple effects are quite different from the main effects (p. 239-240).[52] Neyman ends warning "what curious answers may be given…if the number of replicates is small and if Nature chooses to be frivolous and not behave as the experimenter expects her to do." (p. 241).[52]

In response to Neyman, Yates[53] says "it has never been claimed that the simple effects are identical with the main effects". He suggests "to assume this and to base recommendations on such an assumption implies knowledge other than that provided by the experiment" (p. 246). Yates[53] claims "all arguments which attempt to show that factorial design is ineffective or misleading when interactions exist contain their own refutation, since it is only possible to assess interactions by experiments of the factorial type." (p. 247). He argues Neyman "has exaggerated the importance of his interaction", noting that interactions are identified over a series of experiments not by relying on a single experiment (p. 247). Yates[22] goes further, suggesting that in agriculture, practices are rarely standardised in the way contemplated by Neyman. He[22] argues that, unless it is known before an experiment, which treatment will be best, “we are forced to survey the whole field”, concluding “the experimenter who confines himself to experiments on single factors, making a guess at the final levels of the other factors, is merely emulating the tactics of an ostrich.” (p. 5). So, from the 1930s, there have been strong views about estimands in factorial trials.

With this in mind, we summarise the discussion within the clinical trials literature. Peto[7] did not require there to be no interactions between treatment factors, stating 2-in-1 factorial trials are efficient if the separate effects ($\theta_{A1}$ and $\theta_{A2}$ or $\theta_{B1}$ and $\theta_{B2}$) are both positive. He argued that otherwise, such trials “point unbiasedly to the complicated truth, while misleading conclusions could well emerge from other designs” (p. 35). In terms of 2-in-1 factorial trials, Stampfer *et al*[10] said “it is the possible complications of an interaction which make many hesitate to use a factorial design” (p. 114). Such complications led Brittain and Wittes[44] to conclude that 2-in-1 factorial trials are efficient “when one expects at most small interactions” (p. 169). This might be true in a 2-in-1 factorial trial, but factorial trials more generally are still efficient when large interactions are expected. Brittain and Wittes[44] state that “in the presence of interactions [a 2-in-1] factorial design will produce a biased estimate for the effect of a therapy administered alone.” (p. 162). For clarity, the impact of interactions on bias and power in a 2-in-1 factorial trial result from a mismatch between the estimand and treatment contrast, not from the trial design. Byar[9] goes on, “the variances of the two- and three-way interaction estimates [that is, $\widetilde{AB}$, $\widetilde{AC}$, $\widetilde{BC}$ and $\widetilde{ABC}$] are four and sixteen times greater than the main effects [he means the separate effects $\tilde{A}$, $\tilde{B}$ and $\tilde{C}$]. This implies that if one is particularly interested in…[the separate] interactions, one should increase the sample size considerably” (p. 57). Hung[54] highlighted the problems with testing for interactions before fitting the additive model, stating that “two-stage testing can be very misleading in that its true significance level can be greater than the desired level” (p. 659). Lubsen and Pocock[55] then argued “it follows that one can never, in the analysis and reporting of a factorial trial, ignore the question of interaction” (p. 586).

By 2000, statements were getting stronger, and a factorial clinical trial was synonymous with a 2-in-1 factorial trial. Green *et al*[45] say “a common approach to sample size and analysis assumes no statistical interactions” (p. 3424). Continuing “It is sometimes stated that factorial designs are useful in part because they provide information on treatment interactions…this advantage is hampered by poor power…the trial must be designed with four times the number of patients" (p. 3425). This is not true for factorial trials in general. Unhelpfully and perhaps unwittingly, Green *et al*[45] conclude "these difficulties may render the factorial design inappropriate for most clinical trials" (p. 3430). This guidance informed a series of reviews of factorial clinical trials.[6,49,56–59] All six reviews found reporting of the rationale for conducting a factorial trial was poor, so the proportion conducted for 2-in-1 purposes remains unclear. Montgomery *et al*[60] state "the prime issue here is [that] the sample size of the trial…is based on the crucial assumption that there is no interaction" as "more plausible interactions would require greatly increased sample sizes" (p. 2) Greatly increased sample sizes are only necessary in 2-in-1 factorial trials. Byth and Gebski[61] then gave a graphical aid and formulae for exploring the loss of power if interactions are present, suggesting these for informing the choice of study design, accounting for interaction.

By 2010, a general no interactions assumption was accepted, and the impact of interactions on bias and power was being reiterated. Montgomery *et al*[49] stated that a "design that enables simultaneous evaluation of multiple interventions, or components of a complex intervention, has an obvious attraction" (p. 2). Wolbers *et al*[62] suggested that the two-arm parallel-group trial design "is the design of choice" in many cases when comparing combination therapy to standard treatment (i.e., $\theta_{A2} = ab - b$) based on power implications (p. 6). Whelan *et al*[63] and Pandis *et al*[64] summarised the considerations respectively in surgical and orthodontic trials. So, while focus was slowly shifting to situations where interactions between treatment factors are plausible, Design of Experiments or MOST literature was not being referenced.

At the same time, guidance was being given for 2-in-1 factorial trials. In 2013, Kahan,[50] building on,[44,54,64] stated that the possibility of an interaction between treatment factors should not be ignored in the analysis as this "can lead to bias and potentially misleading conclusions" (p. 4540). Furthermore, treatment effect estimates from a two-stage analysis were shown to be biased even in the absence of an interaction.[50] Based on possible bias, Kahan[50] recommended that a four-arm parallel-group design is preferred because "This leads to straightforward interpretation of results, is unbiased regardless of the presence of an interaction, and allows investigators to ensure adequate power" (p. 4548). Freidlin and Korn[57] then emphasise a distinction between a factorial *design*, which they regard as a type of multi-arm design, and a factorial *analysis*, which includes only separate effects. They[57] conclude that "The presence of a statistical interaction compromises the ability of a factorial trial to provide rigorous evidence if the trial is designed using a factorial analysis" (p. 5). In this way, bias and power considerations with a 2-in-1 factorial trial were leading to guidance towards calculating sample size based on the standard errors from a dummy-coded maximal model, not those from the dummy-coded additive model.

Kahan *et al*[59] identified 752 individually randomised 2×2 factorial trials published up to March 2018. They summarised guidance, which had become widely established by this time. If treatment factors interact "results from a factorial analysis can be misleading" and a multi-arm analysis is "more appropriate" because this analysis "is valid even when treatments interact" (p. 53). In addition, they suggest that including an interaction into the dummy-coded additive model "is flawed, as it does not maintain efficiency". They state that it is "identical to a multi-arm analysis using a different parameterization". To clarify, the separate interaction $\widetilde{AB} = \widehat{AB}$ not the combined effect of treatments $A$ and $B$, $ab - (1)$. Kahan *et al* conclude by saying "When efficiency is the main aim, it is generally recommended that factorial trials are undertaken only when no interaction is expected" (p. 53).[59] However this only applies to 2-in-1 factorial trials conducted to estimate separate effects, not to all factorial trials. More generally, factorial trials are still efficient and should be used when interactions are expected.

Finally, Kahan *et al*[1] extended the CONSORT 2010 statement for factorial trials. Helpfully they say that disparate reasons for conducting a factorial trial "require different methodology, including sample size calculations and analysis strategies" (p. 2106).[1] They also make three sensible recommendations: i) the purpose for using a factorial trial design should be reported, along with whether an interaction is plausible, ii) which treatments form the main treatment comparisons should be clearly identified, and iii) the interaction estimated, and its precision, should be reported for each main comparison (p. 2106).[1]

## 5. ILLUSTRATIVE EXAMPLE – Kingkaew *et al*[37]

We return to the example introduced in Section 2.3 but simplify it for illustrative purposes. Here, we assume smokers were randomised to receive one of four sets of text messages, with techniques to enhance their capability ($C$) and opportunity ($O$) to stop smoking. There were two stratification factors, intention-to-quit (high, low) and age (up to 30, 30+). The primary outcome was 7-day abstinence (binary) at one-month after randomisation. Of 1571 smokers randomised, 1310 provided primary outcome data and were included in analyses. Overall, 521/1310 (40%) smokers self-reported being abstinent, with 789/1310 (60%) still smoking. The proportions abstinent for the complete cases are summarised in Table 5.

[Insert Table 5 about here]

It can be seen that the proportion of smokers that were abstinent increased by 2% (from 41% to 43%) when text messages to enhance either their capability or opportunity to stop smoking were sent in isolation ($\theta_{C1} = c - (1)$ and $\theta_{O1} = o - (1)$). Yet, when text messages were sent in combination, the proportion of smokers that were abstinent decreased by 12% (from 43% to 31%). That is, $\theta_{C2} = co - o$ and $\theta_{O2} = co - c$ equal -12%. This difference ($\theta_{C1} \neq \theta_{C2}$ and $\theta_{O1} \neq \theta_{O2}$) indicates there is an interaction between capability and opportunity. As the effect of sending text messages to enhance capability was lower in the presence than in the absence of text messages to enhance opportunity (and vice versa), this is an 'antagonistic' interaction.

Three logistic regression models were fitted dummy-coding the treatment variables (see Table 6 below). For clarity, look first at the regression coefficients and standard errors, which are on a linear scale. In the additive model, also referred to as the 'factorial analysis', the intercept is a weighted average of outcomes in the four treatments, for participants with a low intention to quit and aged 30 years or over. The effects of capability and opportunity are marginal effects ($\tilde{C}$ and $\tilde{O}$) averaged across levels of the other factor. These marginal effects indicate that, on average, text messages to enhance capability and opportunity decrease the rate of abstinence, but not significantly at a 2-sided 5% significance level. Adding an interaction into the additive model to give the maximal model, changes the interpretation of the intercept and the effects of capability and opportunity. The intercept is now the outcome of no text messages for smokers with a low intention to quit and aged 30 years or over. Effects of capability and opportunity are now separate effects of sending capability or opportunity text messages in isolation. Their standard errors are higher than those of the marginal effects, indicating the separate effects are estimated with lower precision. The two separate effects indicate that conditional on the other factor being set to absent, text messages to enhance capability and opportunity increase rates of abstinence, but not significantly at a 2-sided 5% significance level. The separate interaction indicates the antagonistic interaction is significant at a 2-sided 5% significance level. It is thus more appropriate to report the maximal model.

[Insert Table 6 about here]

An alternative parameterisation of the dummy-coded maximal model is the so-called multi-arm analysis (see Table 6). The intercept is the same, as are the separate effects of capability and opportunity. Yet, the regression coefficient for the combined effect estimates a separate effect ($\theta_{CO1} = co - (1)$) not the separate interaction ($\widetilde{CO} = (1) + co - c - o$) from the first parameterisation of the maximal model. The standard error for this separate effect is lower than the standard error for the separate interaction, indicating the combined effect is estimated with higher precision.

Two logistic regression models were fitted effect-coding the treatment variables (see Table 6). Again, look first at the regression coefficients and standard errors, which are on a linear scale. In the additive model, the intercept is the simple average of the outcome in the four treatments for participants with a low intention to quit and aged 30 years or over. Regression coefficients for capability and opportunity are half the main effects ($\hat{C}$ and $\hat{O}$) defined by Yates.[21] They are therefore half the size of the regression coefficients for the marginal effects from the corresponding dummy-coded model. Their standard errors are also half the size of standard errors of the regression coefficients for the marginal effects, so the hypothesis tests are the same. Adding the interaction into the effect-coded additive model has no substantive effect on the interpretation or size of the intercept and the main effects. The two main effects and interaction in the maximal model are estimated with equal precision (the standard errors are all 0.058) and on the same scale. The regression coefficient for half the interaction, $\widehat{CO}/2 = \frac{1}{4}\{(co - o) - (c - (1))\}$, is a quarter the size of that for the separate interaction $\widetilde{CO} = (co - o) - (c - (1))$. The standard error for $\widehat{CO}/2$ is a quarter the size of the standard error for $\widetilde{CO}$ as well, so the hypothesis test is the same.

Simple inspection of Table 5 indicates that there is an antagonistic interaction. A statistically significant antagonistic interaction means that under the principle of marginality, and Bailey's[23] approach to decision making, the maximal model should be reported; the main effects cannot be interpreted without the interaction. Under the principle of heredity and a component screening approach,[5,28] neither main effect being significant in its own right may question the need to consider the interaction, indicating the data support a model without any treatment effects. The conclusion is the same, the optimal treatment here is $(C_1, O_1)$, no text messages to enhance capability or opportunity to stop smoking. Thus, in this 2×2 case, the decision about whether to report the multi-arm analysis, the dummy-coded maximal model, or the effect-coded maximal model depends on which estimands are of interest. All the appropriate models recognise that presenting marginal effects alone will not adequately describe the data. Given that the size of the associated treatment effect estimated, and its precision, vary from model to model, the sample size should follow from the estimands that are of interest and the model that will be used to estimate them.

Note that all the separate effects can be predicted from the effect-coded maximal model, but the reverse is not true of main effects and interactions from a dummy-coded maximal model.

The regression coefficient for $\theta_{C1}$, for example, is equal to $\hat{C} - \widehat{CO}$, which is approximately equal to $(2 \times -0.10) - (2 \times -0.14) = 0.08$. The regression coefficient for $\theta_{O1}$ is equal to $\hat{O} - \widehat{CO}$, which is about $(2 \times -0.11) - (2 \times -0.14) = 0.06$. As the regression coefficient for the interaction is $\widehat{CO}/2 = -0.14$, the regression coefficient for the separate interaction $\widetilde{CO}$ is twice the interaction $\widehat{CO}$, which is about $(2 \times -0.28) = -0.56$. Finally, the regression coefficient for the combined effect $\theta_{CO}$ is equal to $\hat{C} + \hat{O}$, which is about $(2 \times -0.10) + (2 \times -0.11) = -0.42$.

The situation becomes more complex in the 2×2×2 case (see Supplementary Materials for the proportions abstinent for the complete cases for each of the eight treatments and the results of the corresponding five logistic regression models). Again, a maximal model or the multi-arm analysis is more appropriate since there is an antagonistic interaction between capability and opportunity. However, as the sample size is fixed at 1310 smokers providing outcome data, the standard errors for the separate effects are larger than in the 2×2 case. Moreover, in the dummy-coded maximal model, the standard error for the three-way separate interaction is larger than the standard errors for the two-way separate interactions. In contrast, the standard errors for the main effects, two-way and three-way interactions from an effect-coded maximal model are the same, not only as each other, but also as they were in the 2×2 case. This shows why a factorial design is efficient for estimating main effects and interactions, especially as the number of treatment factors increases. The hypothesis test for the three-way separate interaction from the dummy-coded maximal model is the same as that for the three-way interaction from the effect-coded maximal model. However, the tests for the three two-way separate interactions and the three two-way interactions differ because they are estimating different underlying parameters. The difference in the underlying parameter estimated by the different interactions means that the specific interaction reported needs to be specified. As before, the separate effects can be predicted from the effect-coded maximal model.

## 6. CONCLUSIONS AND RECOMMENDATIONS

The role of interactions in factorial trials depends on the purpose of the trial. If the purpose of a trial is to compare two or more largely unrelated interventions to a control more efficiently, this efficiency comes from assuming that there are no interactions between treatment factors. However, if the purpose of the trial is to screen out components of a complex intervention, to empirically optimise a complex intervention, or to confirm the optimal complex intervention, efficiency comes from the inclusion of all patients when estimating each of the estimands of interest (i.e., main effects and interactions). Here, the presence of interactions does not affect the statistical power to detect main effects, assuming the sample size is based on estimating the main effects, which now differ from the separate effects. Here also, the statistical power to detect interactions is comparable to the power for detecting main effects if these quantities are anticipated to be of a similar size. Thus, it is essential that clinical trialists carefully consider the rationale for their trial at the design stage, identifying and reporting their estimands of interest. Without this being clear, it is not possible to understand whether the trial design and analysis strategy proposed or used are appropriate.

Blanket statements about factorial trials are unhelpful. Much of the guidance is specific to the $k$-in-1 factorial trial, where there is interest in estimating separate effects (but also the separate interactions to check their assumption of no interactions), not main effects or interactions. As we have seen, Green *et al*[45] conjectured that "[o]ccasionally…two primary questions… may be considered independent questions. More often…the goal at the end…is to recommend for use one of the four treatment arms" (p. 3425). It is therefore possible that the assumption of no interactions in factorial clinical trials has been over-emphasised and that clinical trialists are actually interested in the main effects and interactions. This is an area for future research.

We have focused on the situation where each treatment is equally replicated (i.e., a balanced or orthogonal design). In practice, non-orthogonal designs may be more common (and our illustrative example was not perfectly balanced). In this situation, the comparison of an effect-coded model with a dummy-coded model will be more complicated because the off-diagonal elements of the variance-covariance matrix of the estimators will not be precisely zero in either case. Any correlation between estimators will be markedly lower in the effect-coded model, however. The implication is that the order in which effects are fitted in the model becomes particularly important in the dummy-coded model. Simplification of the model will be more controversial because the estimates of the treatment effects will differ depending on whether the maximal model is fitted or a reduced model. The amount of change will depend on the degree of non-orthogonality in the design. Some non-orthogonality may have little impact, as we have seen. Situations where there is substantial non-orthogonality are also an area for future research.

We have also focused on the situation where each of the treatment factors has only two levels. It is entirely possible to generalise this to treatment factors with three or more levels, using a larger set orthogonal contrasts. For three levels of factor $A$ labelled 1, 2 and 3, for example, the two orthogonal contrasts for the main effects of factor $A$ may be one coded 1/2, 1/2 and –1 (contrasting levels 1 and 2 to level 3) and one coded 1, –1 and 0 (contrasting level 1 to level 2). The overall sample size will increase to become comparable to a 3-arm parallel-group trial, however, when one or more factors have three levels.

Given the extent of statistical input common in modern clinical trials, perhaps now is the time to embrace more complex experimental methods in health and social care research, especially for trials of complex interventions. Building on the MOST research framework for preparing, optimising and evaluating complex interventions, we believe that greater appreciation of methods, which have been long established in the Design of Experiments, could greatly benefit clinical trials. It is now 100 years since Fisher[20] first recommended factorial experiments in agriculture. We believe that it is time for medicine to catch up, given that the difficulties that have purportedly rendered factorial designs inappropriate for most clinical trials [45] are not always relevant.

## References:

1. Kahan BC, Hall SS, Beller EM, Birchenall M, Chan AW, Elbourne D, Little P, Fletcher J, Golub RM, Goulao B, Hopewell S, Islam N, Zwarenstein M, Juszczak E, Montgomery AA. Reporting of Factorial Randomized Trials: Extension of the CONSORT 2010 Statement. JAMA. 2023; 330(21): 2106-2114.
2. Campbell M, Fitzpatrick R, Haines A, Kinmonth AL, Sandercock P, Spiegelhalter D, Tyrer P. Framework for design and evaluation of complex interventions to improve health. British Medical Journal. 2000; 321(7262): 694-696
3. Craig P, Dieppe P, Macintyre S, Mitchie S, Nazareth I, Petticrew M. Developing and evaluating complex interventions: The new Medical Research Council guidance. British Medical Journal. 2008; 337(7676): 979-983
4. Skivington K, Matthews L, Simpson SA, Craig P, Baird J, Blazeby JM, Boyd KA, Graig N, French DP, McIntosh E, Petticrew M, Rycroft-Malone J, White M, Moore L. A new framework for developing and evaluating complex interventions: Update of Medical Research Council guidance. British Medical Journal. 2021; 374: n2061
5. Box GEP, Hunter WG, Hunter JS. Statistics for Experimenters: An introduction to design, data analysis, and model building. New York: Wiley; 1978.
6. McAlister FA, Straus SE, Sackett DL, Altman DG. Analysis and reporting of factorial trials: A systematic review. JAMA. 2003; 289(19): 2545-2553.
7. Peto R. Clinical trial methodology. Biomedicine. 1978; 28 Spec No: 24-36.
8. Chalmers TC, Eckhardt RD, Reynolds WE, Cigarron JG, Denne N, Reifenstein RW, Smith CW, Davidson CS. The treatment of acute infectious hepatitis. Controlled studies on the effect of diet, rest, and physical reconditioning on the acute course of the disease and on the incidence of relapses and residual abnormalities. Journal of Clinical Investigation. 1955; 34: 1163-235.

9. Byar DP. Factorial and reciprocal control designs. Statistics in Medicine. 1990; 9: 55-64.
10. Stampfer MJ, Buring JE, Willett W, Rosner B, Eberlein K, Hennekens CH. The 2x2 factorial design: Its application to a randomized trial of aspirin and carotene in US physicians. Statistics in Medicine. 1985; 4: 111-116.
11. Doll R. Controlled trials testing two or more treatments simultaneously. Journal of the Royal Society of Medicine. 2005; 98: 479–480.
12. Wilson C, Pollock MR, Harris AD. Diet in the treatment of infectious hepatitis. Lancet. 1946; i: 881–883.
13. Wyckoff J, Dubois EF, Woodruff IO. The therapeutic value of digitalis in pneumonia. JAMA. 1930; 95: 1243–1249.
14. Doll R, Pygott F. Factors influencing the rate of healing of gastric ulcers: Admission to hospital, phenobarbitone, and ascorbic acid. Lancet. 1952; i: 171–175.
15. Chalmers TC. A potpourri of RCT topics. Controlled Clinical Trials. 1982; 3: 285-298.
16. Doll R. Clinical trials: retrospect and prospect. Statistics in Medicine. 1982; 1: 337-344.
17. Fisher RA. The Design of Experiments. London: Oliver and Boyd; 1935.
18. Collins LM. Optimization of Behavioral, Biobehavioral and Biomedical Interventions: The multiphase optimization strategy (MOST). New York: Springer; 2018.
19. Collins LM, Kugler KC. Optimization of Behavioral, Biobehavioral and Biomedical Interventions: Advanced topics. New York: Springer; 2018.
20. Fisher RA. The arrangement of field experiments. Journal of the Ministry of Agriculture. 1926; 33: 503-15.
21. Yates F. Complex experiments. Supplement to the Journal of the Royal Statistical Society. 1935; 2(2): 181-223.
22. Yates F. The Design and Analysis of Factorial Experiments (Technical Communication No. 35). Harpenden, England: Imperial Bureau of Soil Science; 1937.
23. Bailey RA. Design of Comparative Experiments. Cambridge: Cambridge University Press; 2008.
24. Cochran WG, Cox GM. Experimental Designs. 2nd ed. New York: Wiley; 1957.
25. Hinkelmann K, Kempthorne O. Design and Analysis of Experiments, Volume 1: Introduction to Experimental Design. 2nd ed. New York: Wiley; 2008.
26. Mead R, Gilmour SG, Mead A. Statistical Principles for the Design of Experiments: Applications to real experiments. Cambridge: Cambridge University Press; 2012.

27. Milliken GA, Johnson DE. Analysis of Messy Data. Volume 1. Designed experiments. 2nd ed. New York: CRC Press; 2009.
28. Wu CFJ, Hamada MS. Experiments: Planning, Analysis and Optimization. 2nd ed. New York: Wiley; 2009.
29. Cheng C-S. Theory of Factorial Design: Single and Multi-Stratum Experiments. New York: Chapman and Hall; 2013.
30. Mukerjee R, Wu CFJ. A Modern Theory of Factorial Designs. New York: Springer; 2006.
31. Collins LM, Murphy SA, Nair VN, Strecher VJ. A strategy for optimizing and evaluating behavioral interventions. Annals of Behavioral Medicine. 2005; 30(1): 65-73.
32. Collins LM, Nahum-Shani I, Guastaferro K, Strayhorn JC, Vanness DJ, Murphy SA. Intervention optimization: A paradigm shift and its potential implications for clinical psychology. Annual Review of Clinical Psychology. 2024; 20: 21-47.
33. Piper ME, Fiore MC, Smith SS, Fraser D, Bolt DM, Collins LM, Mermelstein R, Schlam TR, Cook JW, Jorenby DE, Loh W-Y, Baker TB. Identifying effective intervention components for smoking cessation: A factorial screening experiment. Addiction. 2016; 111: 129–141.
34. Dasgupta T, Pillai NS, Rubin DB. Causal inference from 2k factorial designs by using potential outcomes. Journal of the Royal Statistical Society Series B: Statistical Methodology. 2014; 77(4): 727-753.
35. ISIS-2 (Second International Study of Infarct Survival) Collaborative Group. Randomised trial of intravenous streptokinase, oral aspirin, both, or neither among 17,187 cases of suspected acute myocardial infarction: ISIS-2. Lancet. 1988; 332(8607): 349-360.
36. Wright-Hughes A, Willis TA, Wilson S, Weller A, Lorencatto F, Althaf M, Seymour V, Farrin AJ, Francis J, Brehaut J, Ivers N, Alderson SL, Brown BC, Feltbower RG, Gale CP, Stanworth SJ, Hartley S, Colquhoun H, Presseau J, Walwyn R, Foy R. A randomised fractional factorial screening experiment to predict effective features of audit and feedback. Implementation Science. 2022; 17(1): 34.
37. Kingkaew P, Glidewell E, Walwyn R, Khuntha S, Yunibhand J, Wyatt JC. Enhancing mobile phone text messaging to support smoking cessation with behavioural change techniques: A 2×2×2 factorial randomised controlled trial. 2020 (Unpublished work).
38. Stanworth S, Walwyn R, Grant-Casey J, Hartley S, Moreau L, Lorencatto F, Francis J, Gould N, Swart N, Rowley M, Morris S, Grimshaw J, Farrin A, Foy R for the AFFINITIE Collaborators. Effectiveness of enhanced performance feedback on appropriate use of blood transfusions: A comparison of 2 cluster randomized trials. JAMA Network Open. 2022; 5(2): e220364.

39. Yusuf S, Collins R, Peto R. Why do we need some large, simple randomized trials? Statistics in Medicine. 1984; 3(4): 409-20.
40. Michie S, van Stralen MM, West R. The behaviour change wheel: A new method for characterising and designing behaviour change interventions. Implementation Science. 2011; 23(6): 42.
41. Collins LM, Trail JB, Kugler KC, Baker TB, Piper ME, Mermelstein RJ. Evaluating individual intervention components: Making decisions based on the results of a factorial screening experiment. Translational Behavioral Medicine. 2014; 4(3): 238-251.
42. Strayhorn JC, Cleland CM, Vanness DJ, Wilton L, Gwadz M, Collins LM. Using decision analysis for intervention value efficiency to select optimized interventions in the multiphase optimization strategy. Health Psychology. 2024; 43(2): 89-100.
43. Neuhauser D, Provost SM, Provost LP. It is time to reconsider factorial designs: How Bradford Hill and R. A. Fisher shaped the standard of clinical evidence. Quality Management in Healthcare. 2020; 29(2): 109–122.
44. Brittain E, Wittes J. Factorial designs in clinical trials: The effects of non-compliance and subadditivity. Statistics in Medicine. 1989; 8: 161-171.
45. Green S, Liu P-Y, O'Sullivan J. Factorial design considerations. Journal of Clinical Oncology 2002; 20: 3424-3430.
46. Kahan BC, Morris TP, Goulão B, Carpenter J. Estimands for factorial trials. Statistics in Medicine 2022; 1-12. DOI: 10.1002/sim.9510
47. ICH E9 (R1) addendum on estimands and sensitivity analysis in clinical trials to the guideline on statistical principles for clinical trials https://www.ema.europa.eu/en/documents/scientific-guideline/ich-e9-r1-addendum-estimands-sensitivity-analysis-clinicaltrials-guideline-statistical-principles_en.pdf.
48. Piantadosi S. Clinical Trials: A methodologic perspective. New York: Wiley; 1997.
49. Montgomery AA, Astin MP, Peters TJ. Reporting of factorial trials of complex interventions in community settings: a systematic review. Trials 2011; 12:179
50. Kahan BC. Bias in randomised factorial trials. Statistics in Medicine 2013; 32: 4540-4549.
51. Korn EL, Freidlin B. Non-factorial analyses of two-by-two factorial trial designs. Clinical Trials 2016; 13(6): 651-659.
52. Neyman J. Discussion on complex experiments. Supplement to the Journal of the Royal Statistical Society. 1935; 2(2): 235-241.
53. Yates F. Reply to discussion on complex experiments. Supplement to the Journal of the Royal Statistical Society. 1935; 2(2): 243-247.

54. Hung HMJ. Two-stage tests for studying monotherapy and combination therapy in two-by-two factorial trials. Statistics in Medicine 1993; 12: 645-660.
55. Lubsen J, Pocock SJ. Factorial trials in cardiology: pros and cons. European Heart Journal 1994; 15: 585-588.
56. Mdege ND, Brabyn S, Hewitt C, Richardson R, Torgerson DJ. The 2×2 cluster randomized controlled factorial trial design is mainly used for efficiency and to explore intervention interactions: A systematic review. Journal of Clinical Epidemiology 2014; 67: 1083-1092.
57. Freidlin B, Korn EL. Two-by-two factorial cancer treatment trials: Is sufficient attention being paid to possible interactions? JNCI Journal of the National Cancer Institute 2017; 109(9): djx146.
58. Goulao B, MacLennan G, Ramsay C. The split-plot design was useful for evaluating complex, multilevel interventions, but there is need for improvement in its design and report. Journal of Clinical Epidemiology 2018; 96: 120-125.
59. Kahan BC, Tsui M, Jairath V, Scott AM, Altman DG, Beller E, Elbourne D. Reporting of randomized factorial trials was frequently inadequate. Journal of Clinical Epidemiology 2020; 117: 52-59
60. Montgomery AA, Peters TJ, Little P. Design, analysis and presentation of factorial randomised controlled trials. BMC Medical Research Methodology 2003; 3:26
61. Byth K, Gebski V. Factorial designs: a graphical aid for choosing study designs accounting for interaction. Clinical Trials 2004; 1: 315-325.
62. Wolbers M, Heemskerk D, Chau TTH, Yen NTB, Caws M, Farrar J, Day J. Sample size requirements for separating out the effects of combination treatments: Randomised controlled trials of combination therapy vs. standard treatment compared to factorial designs for patients with tuberculous meningitis. Trials 2011; 12: 26.
63. Whelan DB, Dainty K, Chahal J. Efficient designs: Factorial randomized trials. The Journal of Bone and Joint Surgery 2012; 94: 34-38.
64. Pandis N, Walsh T, Polychronopoulou A, Katsaros C, Eliades T. Factorial designs: an overview with applications to orthodontic clinical trials. European Journal of Orthodontics 2014; 36: 314-320.
65. Ng T-H. The impact of a preliminary test for interaction in a 2×2 factorial trial. Communications in Statistics: Theory and Methods 1994; 23: 435-450.

**Table 1: Differences Between the Two Schools of Thought**

| | **$k$-in-1 Factorial Trials** | **Other Factorial Trials*** |
|---|---|---|
| **Requirement** | Undertaken in the absence of statistical interactions between treatment factors. | Undertaken in the presence or absence of interactions between treatment factors. |
| **Objective** | To evaluate two or more whole interventions (often drugs) in isolation without increasing the sample size from that required for one parallel group trial. | To build, optimise or evaluate a multi-component intervention. Levels of each treatment factor are versions of an intervention component. |
| **Estimands** | Separate effects (i.e., the effect of each whole intervention when administered alone) and separate interactions (the additional effect of combining interventions). | Main effects (for each component, whether more or less can be added with advantage) and interactions (modifications to additive main effects). |
| **Treatment Contrasts** | In a regression analysis, treatment factors can be added as dummy $(0, 1)$ variables to directly estimate separate effects. An additive model, without any interactions, is desirable but only valid if there are no interactions. | In a regression analysis, treatment factors can be added as effect coded $(-1, +1)$ variables to directly estimate main effects and interactions. A maximal model is recommended, including all the interactions. |
| **Properties of Estimators (under equal replication)** | 1. Estimators are correlated.<br>2. Estimators of the separate effects and their standard errors are changed when adding separate interactions into the model.<br>3. The standard errors of the lower order effects are smaller than those of higher order effects.<br>4. Separate effects and separate interactions are estimated with less precision if more treatment factors are added and the sample size is fixed. | 1. Estimators are orthogonal (uncorrelated).<br>2. Estimators of the main effects and their standard errors are unchanged by adding interactions into the model.<br>3. The standard errors of the main effects and interaction are all equal.<br>4. Main effects and interactions are estimated with the same precision if more treatment factors are added into the design and the sample size is fixed. |

* Note: Other factorial trials include factorial screening experiments, factorial optimisation trials and factorial evaluation trials that vary in how explanatory or pragmatic, exploratory or confirmatory, they are designed to be. Screening experiments aim to build multi-component interventions, optimisation trials aim to optimise them, and evaluation trials aim to evaluate them.

**Table 2. Mean Outcomes and Treatment Effects in a 2×2 Factorial Trial**

| | | Treatment Factor A | | Separate Effects | Main Effect and Interaction |
|---|---|---|---|---|---|
| | | Level 1 = −1 (e.g., Absent) | Level 2 = +1 (e.g., Present) | | |
| Treatment Factor B | Level 1 = −1 (e.g., Absent) | $(A_1, B_1) = (-1, -1)$ $(1)$ | $(A_2, B_1) = (+1, -1)$ $a$ | $\theta_{A1} = a - (1)$ | $\hat{A} = (\theta_{A1} + \theta_{A2})/2$ |
| | Level 2 = +1 (e.g., Present) | $(A_1, B_2) = (-1, +1)$ $b$ | $(A_2, B_2) = (+1, +1)$ $ab$ | $\theta_{A2} = ab - b$ | |
| Separate Effects | | $\theta_{B1} = b - (1)$ | $\theta_{B2} = ab - a$ | $\theta_{AB} = \text{ab} - (1)$ | |
| Main Effect and Interaction | | $\hat{B} = (\theta_{B1} + \theta_{B2})/2$ | | | $\widehat{AB} = (\theta_{A2} - \theta_{A1})/2 = (\theta_{B2} - \theta_{B1})/2$ |

**Table 3: Treatments in Standard Order for a $2^3$ (or 2×2×2) Factorial Design**

| | Treatment | Treatment Label | Mean Outcome |
|---|---|---|---|
| 1 | $(-1, -1, -1)$ | $A_1$ and $B_1$ and $C_1$ | $(1)$ |
| 2 | $(+1, -1, -1)$ | $A_2$ and $B_1$ and $C_1$ | $a$ |
| 3 | $(-1, +1, -1)$ | $A_1$ and $B_2$ and $C_1$ | $b$ |
| 4 | $(+1, +1, -1)$ | $A_2$ and $B_2$ and $C_1$ | $ab$ |
| 5 | $(-1, -1, +1)$ | $A_1$ and $B_1$ and $C_2$ | $c$ |
| 6 | $(+1, -1, +1)$ | $A_2$ and $B_1$ and $C_2$ | $ac$ |
| 7 | $(-1, +1, +1)$ | $A_1$ and $B_2$ and $C_2$ | $bc$ |
| 8 | $(+1, +1, +1)$ | $A_2$ and $B_2$ and $C_2$ | $abc$ |

**Table 4. Main Effects and Interactions in a $2^3$ (or 2×2×2) Factorial Experiment**

| **Mean Treatment Outcome** | **Treatment Effect** | | | | | | | |
|---|---|---|---|---|---|---|---|---|
| | Grand Mean | $\hat{A}$ | $\hat{B}$ | $\widehat{AB}$ | $\hat{C}$ | $\widehat{AC}$ | $\widehat{BC}$ | $\widehat{ABC}$ |
| (1) | + | − | − | + | − | + | + | − |
| a | + | + | − | − | − | − | + | + |
| b | + | − | + | − | − | + | − | + |
| ab | + | + | + | + | − | − | − | − |
| c | + | − | − | + | + | − | − | + |
| ac | + | + | − | − | + | + | − | − |
| bc | + | − | + | − | + | − | + | − |
| abc | + | + | + | + | + | + | + | + |

**Table 5. Proportions of Smokers Abstinent (Primary Outcome) for a 2×2 Factorial Trial (based on Kingkaew *et al*[37])**

| | | Capability Text Messages | | Overall |
|---|---|---|---|---|
| | | Absent | Present | |
| **Opportunity Text Messages** | **Absent** | (1) = 133/322 = 41% | *c* = 142/328 = 43% | 275/650 = 42% |
| | **Present** | *o* = 144/336 = 43% | *co* = 102/324 = 31% | 246/660 = 37% |
| **Overall** | | 277/658 = 42% | 244/652 = 37% | 521/1310 = 40% |

**Table 6: Illustrative Results for the Dummy Coded and Effect Coded Models from the 2×2 Factorial Trial (based on Kingkaew *et al*[37])**

| **Parameterisation** | **Dummy Coded (0, 1) Effects** | | | | **Effect Coded (-1, +1) Effects** | | | |
|---|---|---|---|---|---|---|---|---|
| **Term** | **Coef.** | **SE** | **p-value** | **Odds Ratio (95% CI)*** | **Coef.** | **SE** | **p-value** | **Odds Ratio (95% CI)*** |
| **MODEL: ADDITIVE MODEL** | | | | | | | | |
| Intercept | -0.64 | 0.128 | <0.001 | 0.53 (0.41 to 0.68) | -0.83 | 0.098 | <0.001 | 0.43 (0.36 to 0.53) |
| ***Stratification effects*** | | | | | ***Stratification effects*** | | | |
| Intention-to-quit (ITQ): High | 0.83 | 0.117 | <0.001 | 2.29 (1.82 to 2.88) | 0.83 | 0.117 | <0.001 | 2.29 (1.82 to 2.88) |
| Age: Up to 30 | 0.10 | 0.116 | 0.400 | 1.10 (0.88 to 1.39) | 0.10 | 0.116 | 0.400 | 1.10 (0.88 to 1.39) |
| ***Marginal effects*** *($\tilde{C}$ and $\tilde{O}$)* | | | | | ***Main effects*** *($\hat{C}/2$ and $\hat{O}/2$)* | | | |
| Capability (C) | -0.19 | 0.116 | 0.096 | 0.82 (0.65 to 1.03) | -0.10 | 0.058 | 0.096 | 0.91 (0.81 to 1.02) |
| Opportunity (O) | -0.20 | 0.116 | 0.079 | 0.82 (0.65 to 1.02) | -0.10 | 0.058 | 0.079 | 0.90 (0.81 to 1.01) |
| **MODEL: MAXIMAL MODEL** | | | | | | | | |
| Intercept | -0.76 | 0.140 | <0.001 | 0.47 (0.35 to 0.61) | -0.83 | 0.099 | <0.001 | 0.43 (0.36 to 0.53) |
| ***Stratification effects*** | | | | | ***Stratification effects*** | | | |
| Intention-to-quit (ITQ): High | 0.82 | 0.117 | <0.001 | 2.28 (1.81 to 2.87) | 0.82 | 0.117 | <0.001 | 2.28 (1.81 to 2.87) |
| Age: Up to 30 | 0.09 | 0.117 | 0.444 | 1.09 (0.87 to 1.37) | 0.09 | 0.117 | 0.444 | 1.09 (0.87 to 1.37) |
| ***Separate effects*** *($\theta_{C1}$ and $\theta_{O1}$)* | | | | | ***Main effects*** *($\hat{C}/2$ and $\hat{O}/2$)* | | | |
| Capability (C) | 0.07 | 0.163 | 0.650 | 1.08 (0.78 to 1.48) | -0.10 | 0.058 | 0.088 | 0.91 (0.81 to 1.01) |
| Opportunity (O) | 0.06 | 0.162 | 0.711 | 1.06 (0.77 to 1.46) | -0.11 | 0.058 | 0.068 | 0.90 (0.80 to 1.01) |
| ***Separate interaction*** *($\widetilde{CO}$)* | | | | | ***Interaction*** *($\widehat{CO}/2$)* | | | |
| CO | -0.54 | 0.233 | 0.019 | 0.58 (0.37 to 0.91) | -0.14 | 0.058 | 0.019 | 0.87 (0.78 to 0.98) |
| **MODEL: MULTI-ARM MODEL** | | | | | | | | |
| Intercept | -0.76 | 0.140 | <0.001 | 0.47 (0.35 to 0.61) | - | - | - | - |
| ***Stratification effects*** | | | | | | | | |
| Intention-to-quit (ITQ): High | 0.82 | 0.117 | <0.001 | 2.28 (1.81 to 2.87) | - | - | - | - |
| Age: Up to 30 | 0.09 | 0.117 | 0.444 | 1.09 (0.87 to 1.37) | - | - | - | - |
| ***Treatment effects*** *($\theta_{C1}$, $\theta_{O1}$ and $\theta_{CO}$)* | | | | | | | | |
| Capability (C) | 0.07 | 0.163 | 0.650 | 1.08 (0.78 to 1.48) | - | - | - | - |
| Opportunity (O) | 0.06 | 0.162 | 0.711 | 1.06 (0.77 to 1.46) | - | - | - | - |
| CO Combined | -0.41 | 0.168 | 0.015 | 0.66 (0.48 to 0.92) | - | - | - | - |

* Note: This is the odds rather than the odds ratio for the intercept in each model.

**SUPPLEMENTARY MATERIALS**

**Table A1. Proportions of Smokers Abstinent (Primary Outcome) for a 2×2×2 Factorial Trial (based on Kingkaew *et al*[37])**

| | | Capability Text Messages | | Separate Effects |
|---|---|---|---|---|
| | | **Absent** | **Present** | |
| **Motivation Text Messages: Absent** | | | | |
| **Opportunity Text Messages** | **Absent** | (1) = 62/158 = 39% | $c$ = 73/164 = 45% | $c - (1) = 6\%$ |
| | **Present** | $o$ = 70/165 = 42% | $co$ = 54/158 = 34% | $co - o = -8\%$ |
| **Separate Effects** | | $o - (1) = 3\%$ | $co - c = -11\%$ | $co - (1) = -5\%$ |
| **Motivation Text Messages: Present** | | | | |
| **Opportunity Text Messages** | **Absent** | $m$ = 71/164 = 43% | $cm$ = 69/164 = 42% | $cm - m = -1\%$ |
| | **Present** | $om$ = 74/171 = 43% | $com$ = 48/166 = 29% | $com - om = -14\%$ |
| **Separate Effects** | | $om - m = 0\%$ | $com - cm = -13\%$ | $com - (1) = -10\%$ |

**Table A2: Illustrative Results for the Dummy Coded and Effect Coded Models from the 2×2×2 Factorial Trial (based on Kingkaew *et al*[37])**

| **Parameterisation** | **Dummy Coded (0, 1) Effects** | | | | **Effect Coded (-1, +1) Effects** | | | |
|---|---|---|---|---|---|---|---|---|
| **Term** | **Coef.** | **SE** | **p-value** | **Odds Ratio (95% CI)*** | **Coef.** | **SE** | **p-value** | **Odds Ratio (95% CI)*** |
| **MODEL: ADDITIVE MODEL** | | | | | | | | |
| Intercept | -0.60 | 0.140 | <0.001 | 0.55 (0.42 to 0.72) | -0.84 | 0.099 | <0.001 | 0.43 (0.36 to 0.53) |
| ***Stratification effects*** | | | | | ***Stratification effects*** | | | |
| Intention-to-quit (ITQ): High | 0.83 | 0.117 | <0.001 | 2.29 (1.83 to 2.89) | 0.83 | 0.117 | <0.001 | 2.29 (1.83 to 2.89) |
| Age: Up to 30 | 0.10 | 0.116 | 0.397 | 1.10 (0.88 to 1.39) | 0.10 | 0.116 | 0.397 | 1.10 (0.88 to 1.39) |
| ***Marginal effects*** | | | | | ***Main effects*** | | | |
| Capability (C) | -0.19 | 0.116 | 0.095 | 0.82 (0.66 to 1.03) | -0.10 | 0.058 | 0.095 | 0.91 (0.81 to 1.02) |
| Opportunity (O) | -0.20 | 0.116 | 0.079 | 0.82 (0.65 to 1.02) | -0.10 | 0.058 | 0.079 | 0.90 (0.81 to 1.01) |
| Motivation (M) | -0.07 | 0.116 | 0.534 | 0.93 (0.74 to 1.17) | -0.04 | 0.058 | 0.534 | 0.96 (0.86 to 1.08) |
| **MODEL: MAXIMAL MODEL** | | | | | | | | |
| Intercept | -0.82 | 0.183 | <0.001 | 0.44 (0.31 to 0.63) | -0.83 | 0.099 | <0.001 | 0.43 (0.36 to 0.53) |
| ***Stratification effects*** | | | | | ***Stratification effects*** | | | |
| Intention-to-quit (ITQ): High | 0.83 | 0.117 | <0.001 | 2.29 (1.82 to 2.88) | 0.83 | 0.117 | <0.001 | 2.29 (1.82 to 2.88) |
| Age: Up to 30 | 0.09 | 0.117 | 0.453 | 1.09 (0.87 to 1.37) | 0.09 | 0.117 | 0.453 | 1.09 (0.87 to 1.37) |
| ***Separate effects*** | | | | | ***Main effects*** | | | |
| Capability (C) | 0.18 | 0.232 | 0.434 | 1.20 (0.76 to 1.89) | -0.10 | 0.058 | 0.089 | 0.91 (0.81 to 1.02) |
| Opportunity (O) | 0.12 | 0.232 | 0.600 | 1.13 (0.72 to 1.78) | -0.11 | 0.058 | 0.069 | 0.90 (0.80 to 1.01) |
| Motivation (M) | 0.11 | 0.232 | 0.649 | 1.11 (0.71 to 1.75) | -0.04 | 0.058 | 0.506 | 0.96 (0.86 to 1.08) |
| ***Separate interactions*** | | | | | ***Interactions*** | | | |
| CO | -0.51 | 0.330 | 0.121 | 0.60 (0.31 to 1.14) | -0.14 | 0.058 | 0.019 | 0.87 (0.78 to 0.98) |
| CM | -0.21 | 0.326 | 0.515 | 0.81 (0.43 to 1.53) | -0.06 | 0.058 | 0.291 | 0.94 (0.84 to 1.05) |
| OM | -0.12 | 0.324 | 0.711 | 0.87 (0.47 to 1.67) | -0.04 | 0.058 | 0.508 | 0.96 (0.86 to 1.08) |
| COM | -0.07 | 0.466 | 0.884 | 0.93 (0.37 to 2.33) | -0.01 | 0.058 | 0.884 | 0.99 (0.88 to 1.11) |
| **MODEL: MULTI-ARM MODEL** | | | | | | | | |
| Intercept | -0.82 | 0.183 | <0.001 | 0.44 (0.31 to 0.63) | - | - | - | - |
| ***Stratification effects*** | | | | | | | | |
| Intention-to-quit (ITQ): High | 0.83 | 0.117 | <0.001 | 2.29 (1.82 to 2.88) | - | - | - | - |
| Age: Up to 30 | 0.09 | 0.117 | 0.453 | 1.09 (0.87 to 1.37) | - | - | - | - |
| ***Treatment effects*** | | | | | | | | |
| Capability (C) | 0.18 | 0.232 | 0.434 | 1.20 (0.76 to 1.89) | - | - | - | - |
| Opportunity (O) | 0.12 | 0.232 | 0.600 | 1.13 (0.72 to 1.78) | - | - | - | - |
| Motivation (M) | 0.11 | 0.232 | 0.649 | 1.11 (0.71 to 1.75) | - | - | - | - |
| CO Combined | -0.21 | 0.238 | 0.379 | 0.81 (0.51 to 1.29) | - | - | - | - |
| CM Combined | 0.07 | 0.232 | 0.747 | 1.08 (0.68 to 1.70) | - | - | - | - |
| OM Combined | 0.11 | 0.230 | 0.642 | 1.11 (0.71 to 1.75) | - | - | - | - |
| COM Combined | -0.50 | 0.242 | 0.037 | 0.60 (0.37 to 0.97) | - | - | - | - |

* Note: This is the odds rather than the odds ratio for the intercept in each model

**Table A3: Intended Results for Kingkaew *et al*[37] from the Effect Coded and Dummy Coded Maximal Models**

| **Parameterisation** | **Effect Coded (-1, +1) Effects** | | | | **Dummy Coded (0, 1) Effects** | | | |
|---|---|---|---|---|---|---|---|---|
| **Term** | **Coef.** | **SE** | **p-value** | **Odds Ratio (95% CI)*** | **Coef.** | **SE** | **p-value** | **Odds Ratio (95% CI)*** |
| Intercept | -0.38 | 0.059 | <0.001 | 0.68 (0.61 to 0.77) | -0.93 | 0.231 | <0.001 | 0.39 (0.25 to 0.61) |
| ***Stratification effects*** | | | | | ***Stratification effects*** | | | |
| Intention-to-quit (ITQ): High | 0.41 | 0.059 | <0.001 | 1.51 (1.34 to 1.69) | 1.08 | 0.340 | 0.001 | 2.96 (1.53 to 5.81) |
| Age: Up to 30 | 0.05 | 0.059 | 0.436 | 1.05 (0.93 to 1.17) | 0.09 | 0.117 | 0.436 | 1.10 (0.87 to 1.38) |
| ***Main effects*** | | | | | ***Separate effects*** | | | |
| Capability (C) | -0.11 | 0.059 | 0.067 | 0.90 (0.80 to 1.01) | 0.24 | 0.313 | 0.448 | 1.27 (0.69 to 2.35) |
| Opportunity (O) | -0.12 | 0.059 | 0.045 | 0.89 (0.79 to 1.00) | 0.25 | 0.311 | 0.419 | 1.29 (0.70 to 2.38) |
| Motivation (M) | -0.03 | 0.059 | 0.588 | 0.97 (0.86 to 1.09) | 0.30 | 0.315 | 0.334 | 1.36 (0.73 to 2.53) |
| ***Interactions*** | | | | | ***Separate interactions*** | | | |
| CO | -0.15 | 0.059 | 0.013 | 0.86 (0.77 to 0.97) | -0.29 | 0.436 | 0.500 | 0.75 (0.32 to 1.75) |
| CM | -0.05 | 0.059 | 0.386 | 0.95 (0.85 to 1.07) | -0.45 | 0.444 | 0.315 | 0.64 (0.27 to 1.53) |
| OM | -0.03 | 0.059 | 0.656 | 0.97 (0.87 to 1.09) | -0.44 | 0.441 | 0.322 | 0.65 (0.27 to 1.53) |
| COM | 0.00 | 0.059 | 0.952 | 1.00 (0.89 to 1.12) | -0.08 | 0.633 | 0.903 | 0.93 (0.27 to 3.20) |
| C and HighITQ | -0.03 | 0.059 | 0.587 | 0.97 (0.86 to 1.09) | -0.13 | 0.473 | 0.778 | 0.88 (0.35 to 2.21) |
| O and HighITQ | -0.05 | 0.059 | 0.385 | 0.95 (0.85 to 1.07) | -0.30 | 0.469 | 0.522 | 0.74 (0.29 to 1.86) |
| M and HighITQ | 0.05 | 0.059 | 0.431 | 1.05 (0.93 to 1.18) | -0.44 | 0.468 | 0.345 | 0.64 (0.26 to 1.61) |
| CO and HighITQ | -0.06 | 0.059 | 0.287 | 0.94 (0.84 to 1.05) | -0.55 | 0.669 | 0.411 | 0.58 (0.15 to 2.14) |
| CM and HighITQ | 0.07 | 0.059 | 0.233 | 1.07 (0.96 to 1.20) | 0.51 | 0.659 | 0.437 | 1.67 (0.46 to 6.09) |
| OM and HighITQ | 0.09 | 0.059 | 0.115 | 1.10 (0.98 to 1.23) | 0.69 | 0.653 | 0.288 | 2.00 (0.56 to 7.20) |
| COM and HighITQ | 0.01 | 0.059 | 0.918 | 1.01 (0.90 to 1.13) | 0.10 | 0.942 | 0.918 | 1.10 (0.17 to 6.99) |

* Note: This is the odds rather than the odds ratio for the intercept in each model